\pdfoutput=1
\documentclass[usenatbib,useAMS,onecolumn]{mnras}
\usepackage[dvipdfmx]{graphicx}
\usepackage[dvipdfmx]{color}
\usepackage{amsmath,amssymb}
\usepackage{epstopdf}
\usepackage{times}
\usepackage{url}
\usepackage{bm}
\usepackage{xcolor}

\newcommand{\beqa}{\begin{eqnarray}}

\newcommand{\eeqa}{\end{eqnarray}}

\newcommand{\simgt}{\lower.5ex\hbox{$\; \buildrel > \over \sim \;$}}
\newcommand{\simlt}{\lower.5ex\hbox{$\; \buildrel < \over \sim \;$}}

\newcommand{\change}[1]{{\color{black} #1}}

\def\ave#1{\left\langle #1 \right\rangle}
\def\widebar{\accentset{{\cc@style\underline{\mskip10mu}}}}

\newcommand{\deltam}{\delta_{\mathrm{m}}}

\newcommand{\bx}{{\bf x}}

\newcommand{\br}{{\bf r}}

\newcommand{\dr}{\mathrm{d}}

\newcommand{\dsigma}{\Delta\Sigma}
\newcommand{\hdsigma}{\widehat{\dsigma}}

\RequirePackage{lineno}

\def\apjl{{ApJ}}                
\def\apjs{{ApJS}}               
\def\ao{{Appl.~Opt.}}           
\def\aap{{A\&A}}                

\title[Robust covariance estimation of galaxy-galaxy weak
lensing]{Robust covariance estimation of
galaxy-galaxy weak lensing:
validation and limitation of jackknife covariance
}

\author[M. Shirasaki et al.]
{Masato Shirasaki$^{1}$\thanks{E-mail: masato.shirasaki@nao.ac.jp},
Masahiro Takada$^{2}$,
Hironao Miyatake$^{3,2}$,
Ryuichi Takahashi$^{4}$,
\newauthor
Takashi Hamana$^{1}$,
Takahiro Nishimichi,$^{2, 5}$
and
Ryoma Murata$^{2,6}$
\\
$^{1}$National Astronomical Observatory of Japan, 
Mitaka, Tokyo 181-8588, Japan \\
$^{2}$Kavli Institute for the Physics and Mathematics of the Universe
(WPI),
The University of Tokyo Institutes for Advanced\\
 Study (UTIAS), The University of Tokyo, 5-1-5 Kashiwanoha, Kashiwa-shi, Chiba, 277-8583, Japan\\
$^{3}$Jet Propulsion Laboratory, California Institute of Technology, Pasadena, CA 91109, USA\\
$^{4}$Faculty of Science and Technology, Hirosaki University, 
3 Bunkyo-cho, Hirosaki, Aomori, 036-8561, Japan\\
$^{5}$CREST, JST, 4-1-8 Honcho, Kawaguchi, Saitama, 332-0012, Japan\\
$^{6}$Department of Physics, School of Science, The University of
Tokyo, 7-3-1 Hongo, Bunkyo, Tokyo 113-0033, Japan
}

\begin{document}

\date{}

\pagerange{\pageref{firstpage}--\pageref{lastpage}} \pubyear{2016}

\maketitle

\label{firstpage}

\begin{abstract}
We develop a method to simulate galaxy-galaxy weak lensing by
utilizing all-sky, light-cone simulations and their inherent halo
catalogs.  Using the mock catalog to study the error covariance matrix
of galaxy-galaxy weak lensing, we compare the full covariance with the
``jackknife'' (JK) covariance, the method often used in the literature
that estimates the covariance from the resamples of the data itself.  
\change{
We show that there exists the variation of JK covariance over realizations of mock lensing measurements, while the average JK covariance over mocks
can give a reasonably accurate estimation of the true covariance
up to separations comparable with the size of JK subregion.
The scatter in JK covariances is found to be $\sim$10\%
after we subtract the lensing measurement around random points.
}
However, 
the JK method tends to underestimate the covariance at the larger
separations, more increasingly for a survey with a higher number density
of source galaxies.  We apply our method to the the Sloan Digital Sky
Survey (SDSS) data, and  show that the 48 mock SDSS catalogs nicely
reproduce the signals and the JK covariance measured from the real data.
We then argue that the use of the accurate covariance, compared to the
JK covariance, allows us to use the lensing signals at large scales
 beyond a size of the JK subregion, which contains cleaner cosmological
 information in the linear regime.
\end{abstract}


\section{INTRODUCTION}

Cross-correlation of large-scale structure (LSS) tracers, galaxies or
clusters, with shapes of background galaxies, referred as to
galaxy-galaxy weak lensing or stacked lensing, offers a unique means of
measuring the average total matter distribution around the foreground
objects at lens redshift \citep[][]{1996ApJ...466..623B,
1998ApJ...503..531H,Fischeretal:00,
2002MNRAS.335..311G,2004ApJ...606...67H,2006MNRAS.368..715M,2007arXiv0709.1159J,2013ApJ...769L..35O,
2013MNRAS.431.1439G, 2014MNRAS.437.2111V}. In particular, by combining
the weak lensing and auto-clustering correlation of the same foreground
tracers, one can recover the underlying matter clustering and then
constrain cosmology by breaking the degeneracy with galaxy bias
uncertainty
\citep[e.g.,][]{Seljaketal:05,2009MNRAS.394..929C,2013MNRAS.432.1544M,2015ApJ...806....2M,2016arXiv160407871K}.
Combining different probes of LSS and cosmic microwave background (CMB)
will become a standard strategy aimed at achieving the full potential of
ongoing and upcoming wide-area galaxy surveys for addressing the
fundamental physics with the cosmological observables
\citep[e.g.,][]{OguriTakada:11,Schaanetal:16}. The surveys include the
Baryon Oscillation Spectroscopic Survey (BOSS), the Dark Energy
Survey (DES)\footnote{\url{http://hsc.mtk.nao.ac.jp/ssp/}}, the Kilo-Degree
Survey (KiDS)\footnote{\url{http://kids.strw.leidenuniv.nl}}, the Subaru
Hyper Suprime-Cam (HSC)
survey\footnote{\url{http://hsc.mtk.nao.ac.jp/ssp/}}, the Dark Energy
Spectroscopic Instrument (DESI)\footnote{\url{http://desi.lbl.gov}}, and
the
Subaru Prime Focus Spectrograph (PFS) survey \citep{Takadaetal:14} in
coming 5 years, and ultimately the
Large Synoptic Survey Telescope
(LSST)\footnote{\url{https://www.lsst.org}},
Euclid\footnote{\url{http://sci.esa.int/euclid/}} and
WFIRST\footnote{\url{http://wfirst.gsfc.nasa.gov}} within a coming 10 year
time scale.

In order to properly extract cosmological information from a given
survey, it is important to understand the statistical properties of LSS
probes that arise from the properties of the underlying matter
distribution. The statistical precision of galaxy-galaxy weak lensing
measurements is determined by the covariance matrix that itself contains
two contributions: the measurement noise and sample variance caused by
an incomplete sampling of the fluctuations due to a finite size of a
survey volume.
An accurate estimation of the covariance is becoming a
challenging issue for upcoming wide-area galaxy surveys
\citep{Hartlap2007,DodelsonSchneider:13,Tayloretal:13}, especially if
the dimension of data vector is large, e.g. when combining different LSS
probes.

Even though the initial density field is nearly Gaussian, the sample
variance of LSS probes gets substantial non-Gaussian contributions from
the nonlinear evolution of large-scale structure
\citep[e.g.,][]{Scoccimarro:99,CoorayHu:01}. Since the different Fourier
modes are no longer independent but rather are correlated with each
other in the weakly or deeply nonlinear regime, it is important to
accurately model the non-Gaussian contribution to the sample variance.
For this, it is now recognized that the super-sample covariance (SSC),
which arises from mode-couplings of sub-survey (observable) modes with
super-survey (unobservable) modes comparable with or greater than the
size of a survey volume, is the largest non-Gaussian contribution to the sample
variance for the cluster counts, the matter power spectrum and the
cosmic shear statistics \citep{2003ApJ...584..702H,Hamiltonetal:06,
TakadaBridle:07, TakadaJain:09, Sato2009, Sato2011,
Takahashietal:09,
Kayoetal:13,KayoTakada:13,
2013PhRvD..87l3504T,TakadaSpergel:14, Schaanetal:14, Takahashietal:14,
Lietal:14b,2015MNRAS.453.3043S,KrauseEifler:16,Mohammedetal:16}. In
particular \citet{2013PhRvD..87l3504T} developed a unified approach to
describing the SSC effect in terms of the response of a given observable
to a background mode modeling the super-survey mode \citep[also
see][]{Lietal:14a,Lietal:14b}. Several cosmic shear
measurements have taken into account the SSC contribution in the
cosmological analysis
\citep{Beckeretal:15,HarnoisDerapsetal:16,Hildebrandtetal:16}.

However, the covariance matrix of galaxy-galaxy weak lensing has not
been fully studied. A commonly-used approach to estimating the
covariance is the jackknife method, which is a well-known method in the
fields of statistics or data analysis \citep[][]{Efron:82} \citep[also
see][for the pioneer cosmological applications]{Bothunetal:83}. The JK
method gives an internal estimator of the errors, i.e. estimating the
covariance from the resamples of subdivided copies of the data itself
\citep[e.g.,][for the use of JK covariance for the galaxy-galaxy weak
lensing]{2013MNRAS.432.1544M,Cacciatoetal:14,Couponetal:15,2016PhRvL.116d1301M,Clampittet:16}.
The advantage of the JK method is that it can account for various
observational effects inherent in the data, such as inhomogeneities in
the depth and measurements across the area. However, the drawback is
that the JK covariance is generally noisy because it is estimated from
one particular realization of the fluctuations, i.e. data itself. For
the same reason, the JK covariance can be unstable or even singular
especially if the dimension of data vector is comparable with 
the number of JK resamples, which could often happen if dividing the
data into different bins of the physical quantities or combining
different probes. Furthermore, \citet{2016arXiv161100752S}
recently showed that the use of the random catalog of lensing galaxies
(or clusters) is important for the covariance estimation of
galaxy-galaxy lensing, differently from the cosmic shear covariance.
What does the JK method really estimate? What are the limitations? These
questions have not been fully addressed yet \citep[also see][for a study
based on the similar motivation for the clustering correlation
function]{Norbergetal:09}.  In particular it is not clear whether the JK
method can capture the SSC contribution in the galaxy-galaxy weak
lensing measurements.

Hence the purpose of this paper is to study the covariance matrix of
galaxy-galaxy weak lensing. To do this, we develop a method to construct
a mock catalog of the galaxy-galaxy weak lensing measurements by fully
utilizing a set of full-sky, light-cone simulations containing the
lensing fields in multiple source planes as well as dark matter and halo
distributions in multiple lens planes \citep{2015MNRAS.453.3043S} (also
see Takahashi et al. in preparation). In order to properly simulate
properties of source galaxies, we populate a ``real'' catalog of source
galaxies into the light-cone simulation realization \citep[also
see][]{Shirasaki2014}. In this way we can take into account the observed
characteristics of source galaxies (their angular distributions,
redshifts, intrinsic ellipticities, and the survey geometry). By
identifying halos that are considered to host galaxies and clusters as
for foreground tracers, based on a prescription such as the halo
occupation distribution and the halo mass-cluster observable scaling
relation, we can make hypothetical galaxy-galaxy weak lensing
measurements from the mock catalog.  With the help of such mock
catalogs, we will study the importance of the SSC contribution to the
sample variance in the galaxy-galaxy weak lensing as well as the
limitations of the JK method.
We then apply the developed method to the Sloan Digital Sky Survey
(SDSS) data: the catalog of source galaxies \citep{2013MNRAS.432.1544M}
and the lens samples of redMaPPer clusters \citep{2014ApJ...785..104R}
and the luminous red galaxies \citep{2001AJ....122.2267E}. We will also
discuss how the use of accurate covariance could improve cosmological
constraints, compared to the JK method.

The paper is organized as follows.  Section~\ref{sec:stack} summarizes
basics of galaxy-galaxy weak lensing, the estimator of the signal, and
the different covariance estimators.  Section~\ref{sec:sim} describes
the details of $N$-body simulations, halo catalogs and ray-tracing
simulations used in this paper, and describes the details of our method
for generating mock catalogs of galaxy-galaxy weak lensing.
Section~\ref{sec:result} presents the main results including detailed
comparison of the different covariance estimators as well as the
application to the real SDSS data.  We discuss the results in
Section~\ref{sec:conclusions}.

\section{Galaxy-galaxy weak lensing}\label{sec:stack}

\subsection{Basics}

Cross-correlating positions of large-scale structure tracers, galaxies
or galaxy clusters, with shapes of background galaxies, known as
galaxy-galaxy weak lensing or stacked cluster lensing, measures the
average mass distribution around the foreground objects:
\beqa
\Sigma(R)=\bar{\rho}_{\rm m0}\int\!\mathrm{d}\chi\,
\left[1+
\xi_{\rm lm}\left(
\sqrt{\chi^2+R^2}\right)\right],
\label{eq:sigma}
\eeqa
where $\bar{\rho}_{\rm m0}$ is the mean matter density of the universe
today, $\chi$ is the comoving radial separation, and $R$ is the
projected separation. $\xi_{\rm lm}(r)$ is the three-dimensional
cross-correlation of matter and lensing objects:
\beqa
\xi_{\rm lm}(r)=\langle \delta_{\rm l}(\bx) \delta_{\rm m}(\bx+\br)\rangle,
\eeqa
where $\delta_{\rm l}(\bx)$ is the number density fluctuation field of
lensing objects and $\delta_{\rm m}(\bx)$ is the mass density
fluctuation field.  The surface mass density profile is related to the
observable quantity of weak lensing, the tangential shear distortion
$\gamma_+$ of the shapes of background galaxies, via
\begin{eqnarray}
 \Delta\Sigma(R)=\Sigma_{\rm cr}\gamma_+(R)=\overline{\Sigma}(R)-\Sigma(R),
\end{eqnarray}
where $\Sigma_{\rm cr}(z_{\rm l},z_{\rm s})$ is the lensing efficiency function,
defined for a system of lens objects and sources at redshifts $z_{\rm l}$
and $z_{s}$, respectively, for a flat universe as
\beqa
 \Sigma_{\rm cr}(z_{\rm l},z_{\rm s})^{-1}=4\pi G(1+z_l)\chi(z_{\rm
  l})\left[1-\frac{\chi(z_{\rm l})}{\chi(z_{\rm s})}\right],
\eeqa
where $\chi(z)$ is the comoving radial distance to redshift $z$.
The subscripts
``l" or ``s" stand for lens or source respectively.
We will often call $\Delta\Sigma$ the
excess surface mass density profile. In practice, we need to properly
take into account the redshift distributions of lensing objects and
source galaxies in the galaxy-galaxy weak lensing measurements. 

Taking into account the redshift width of lensing objects and the
connection of lensing objects with the underlying halo distribution, the
lensing profile $\Delta\Sigma(R)$ can also be expressed as
\beqa
 \Delta\Sigma(R)=\frac{\bar{\rho}_{\rm m0}}{\bar{N}_{\rm l}}
  \int\!\mathrm{d}\chi~f_{\rm l}(\chi;z_{\rm l})~\int\!\mathrm{d}M~
  \frac{\mathrm{d}n}{\mathrm{d}M}S_{\rm l}(M;\chi)
  \int\!\frac{k\mathrm{d}k}{2\pi}P_{\rm hm}(k;M, \chi)J_2(kR),
  \label{eq:dsigma_theory}
\eeqa
where
\beqa
 \bar{N}_{\rm l}=\int\!\mathrm{d}\chi~f(\chi;z_{\rm l})
  \int\!\mathrm{d}M~\frac{\mathrm{d}n}{\mathrm{d}M}S_{\rm l}(M;\chi).
\eeqa
Here $\bar{N}_{\rm l}$ is the average projected number density of
lensing objects in units of $[{\rm Mpc}^{-2}]$, $f_{\rm l}(\chi;z_{\rm
l})$ is the radial selection function for lensing objects around $z_{\rm
l}$, $\mathrm{d}n/\mathrm{d}M$ is the halo mass function, $S_{\rm l}(M)$
is the selection function of halos that describes how to populate
lensing objects into halos of mass $M$ and at redshift $z=z(\chi)$ in an
average sense (e.g. the halo occupation distribution), $P_{\rm
hm}(k;M,\chi)$ is the cross-power spectrum between distributions of
matter and halos with $M$ and $z$, and $J_2(x)$ is the second-order
Bessel function. The above equation implicitly assumes that each lensing
object is at the center of its host halo.  If off-centered lensing
objects are included in the analysis, the equation needs to be modified,
according to the method in \cite{OguriTakada:11} \citep[also
see][]{Hikageetal:13,2013MNRAS.435.2345H,2015ApJ...806....2M}.

\subsection{Estimator of galaxy-galaxy weak lensing}

In an actual observation, the galaxy-galaxy weak lensing has to be
estimated from the observed ellipticities of source galaxies.  In this
paper, we employ the estimator of $\Delta \Sigma$ in
\citet{2013MNRAS.432.1544M};
\beqa
 \hdsigma(R)\equiv \frac{1}{2{\cal R}}
  \left[
  \frac{1}{N_{\rm ls}}
  \left.
   \sum_{\rm l,s}~\Sigma_{\rm
   cr}(z_{\rm l},z_{\rm s})w_{\rm ls}
   \epsilon_+(\btheta_s)\right|_{R=\chi_{\rm l}\Delta\theta}
-
  \frac{1}{N_{\rm rs}}
  \left.
   \sum_{\rm r,s}~\Sigma_{\rm
   cr}(z_{\rm r},z_{\rm s})w_{\rm rs}
   \epsilon_+(\btheta_s)\right|_{R=\chi_{\rm r}\Delta\theta}
  \right],
\label{eq:est_delta_Sigma}
\eeqa
where the subscript ``r'' denotes the random catalog of lensing
objects, the projected (comoving) radius $R$ are estimated from the
observed angular separation for lens- or random-source pair, $\Delta
\theta$, e.g., via $R = \chi_{\rm l}\Delta \theta$, $\epsilon_{+}$
represents the tangential component of observed ellipticity of source
with respect to the center of lensing object or random point, and
$w_{\rm ls}$ or $w_{\rm rs}$ is the weight. The summation runs over all
the pairs of sources and lenses (or randoms) separated by the projected
radius $R$ to within a given bin width.  ${\cal R}$ is the responsivity
that is needed for conversion of ellipticity to lensing shear
$\gamma_+$, due to the definition of ellipticity used in this paper
\citep{BernsteinJarvis:02}. Note that the random catalog is built
so as to follow the survey geometry (boundary) and the survey depth as
well as to resemble properties of lensing objects such as the redshift
distribution and the halo mass distribution.  
The above estimator
is defined in analogy with \citet{LandySzalay:93}. 
The stacked lensing around random points is generally non-zero at large
radii due to a violation of periodic boundary conditions in a general
survey geometry. The non-lensing ($B$-mode) signal is also non-zero at
the large radii for the same reason. The subtraction of the stacked
lensing around random points corrects for these effects. Furthermore, as
pointed out in \citet{2016arXiv161100752S}, the use of the random
catalogs is important for the covariance estimation of galaxy-galaxy
weak lensing.

Again following \citet{2013MNRAS.432.1544M}, we employ the weight that
is motivated by the inverse variance weighting, assuming that the
measurement error is dominated by the intrinsic ellipticity on
individual galaxy basis:
\beqa
w_{\rm ls}=\frac{1}{\Sigma_{\rm
 cr}(z_{\rm l},z_{\rm s})^2\left(\sigma_{\epsilon}^2+\sigma_{\rm SN}^2\right)}, 
\label{eq:w_ls}
\eeqa
where $\sigma_{\rm SN}$ is the rms intrinsic ellipticity and
$\sigma_{\epsilon}$ is the measurement error of galaxy ellipticity.
The factor of $\Sigma_{\rm cr}^{-2}$ accounts for downweighting pairs
that are close in redshift because the lensing efficiency is suboptimal.
In reality, we might need to make further corrections of other residual
systematic errors at each radial bin e.g., a multiplicative bias in
shape measurements and an inclusion of unlensed galaxies into the source
galaxy sample \citep[see][for details]{2005MNRAS.361.1287M}.  
The weight for the random points, $w_{\rm rs}$, is defined in the similar
manner to the above equation, since each random point has assigned
ellipticity, measurement error and redshift that mimic the distributions
of lensing galaxies or clusters.

Let us give an intuitive expression of the galaxy-galaxy lensing that
might be useful for the following discussion. Suppose that the observed
ellipticity of a source galaxy is given as a sum of the intrinsic shape
and the lensing distortion effect due to the intervening matter
distribution in the weak lensing regime:
\beqa
 \epsilon_+ \simeq \epsilon^{\rm int}_+ + \gamma_+(\btheta_{\rm s})
\eeqa
The lensing field $\gamma_+$ is equivalent to the lensing convergence
field and is expressed in terms of the mass density fluctuation field as
\begin{eqnarray}
 \gamma_+(\btheta_{\rm s})\leftrightarrow \kappa(\btheta_{\rm s})&=&
 \bar{\rho}_{\rm m0} \int^{\chi_{\rm s}}_0\!\dr \chi~ \Sigma_{\rm
  cr}(\chi,\chi_{\rm s})^{-1}\deltam(\chi,\chi\btheta_{\rm s}) \nonumber\\
&\simeq &\bar{\rho}_{\rm m0}\Sigma_{\rm cr}(z_{\rm l},z_{\rm s})^{-1}\int_{{\rm around}
~ z_{\rm l}}\!\dr\chi~ \deltam(\chi,\chi\btheta_{\rm s})
+\bar{\rho}_{\rm m0} \int_{z\ne z_{\rm l}}\!\dr \chi~ \Sigma_{\rm
cr}(\chi,\chi_{\rm s})^{-1}\deltam(\chi,\chi\btheta_{\rm s}).
\label{eq:shear_decompose}
\end{eqnarray}
In the last line on the r.h.s. we intended to explicitly show that the
lensing effect on a source galaxy in the direction $\btheta_s$ and at
redshift $z_s$ arises from the mass distribution in the lens planes
around redshift $z_{\rm l}$ as well as all the mass distribution along
the line of sight. This equation shows that, by cross-correlating the
shapes of background galaxies with positions of lensing objects in a
particular range of lens redshift, we can probe the average mass
distribution in the lens planes (the first term on the r.h.s.), while
the matter distribution at different redshifts causes statistical
errors (the 2nd term).

\subsection{Covariance matrix}
\label{sec:cov}

The covariance matrix characterizes statistical errors in measurements
of the stacked lensing profile. The statistical errors arise from the
shape noise of source galaxies, the shot noise arising due to a finite
number of lens-source pairs, and the sample variance of large-scale
structure at lens redshift and along the line of sight.

As discussed in \citet{2013PhRvD..87l3504T}, the covariance matrix is
generally expressed by a sum of the three distinct contributions, the
Gaussian covariance, the non-Gaussian (NG) covariance and the
super-sample covariance (SSC), respectively:
\begin{eqnarray}
 {\bf C}_{ij}&\equiv & \ave{\hdsigma(R_i)\hdsigma(R_j)}-\ave{\hdsigma(R_i)}\ave{\hdsigma(R_j)}\nonumber\\
  &=&{\bf C}^{\rm Gauss}_{ij}+{\bf C}^{\rm NG0}_{ij}+{\bf
   C}^{\rm SSC}_{ij}.
   \label{eq:cov_def}
\end{eqnarray}
Exactly speaking, the SSC contribution is a part of the non-Gaussian
covariance, but we separated the SSC because it arises from super-survey
modes, the mass density fluctuation field whose length scales are
comparable with or greater than a size of the survey region. With this
decomposition, we impose that the non-Gaussian term ${\bf C}^{\rm NG0}$
arises from the trispectrum of sub-survey modes.  The Gaussian and
non-Gaussian terms, ${\bf C}^{\rm G}$ and ${\bf C}^{\rm NG0}$, scale
with survey area ($\Omega_s$) as ${\bf C}^{\rm Gauss}, {\bf C}^{\rm NG0}
\propto 1/\Omega_s$. The SSC amplitude scales with the variance of the
coherent density mode in the survey window at each lens redshift,
$\sigma_{{\rm b}, z_{\rm l}}^2\equiv \langle\delta_{{\rm b},z_{\rm
l}}^2\rangle$, where $\delta_{{\rm b}; z_{\rm
l}}=\int\!\mathrm{d}^3\bx~\delta_{\rm m}(\bx)W(\bx;z_{\rm l})$ and
$W(\bx;z_{\rm l})$ is the survey window around the lens slice at $z_{\rm
l}$.  For a fixed area, the amplitude also depends on the survey
geometry \citep{2013PhRvD..87l3504T,Takahashietal:14}.  Thus the SSC
causes a complication in the covariance estimation: since the
super-survey modes are not a direct observable from a given survey, one
has to model the contribution theoretically, either based on the
analytic method or the mock catalogs based on simulations properly
including the super-survey modes. This paper employs the latter
approach.

The goal of this paper is to develop a method of accurately estimating
the covariance matrix for a given survey as well as to develop a
physical understanding of how each term in equation~(\ref{eq:cov_def})
contributes the covariance matrix for ongoing and existing surveys such
as the SDSS dataset. In the literature, several methods for the
covariance estimation have been considered. For example, one way is the
jackknife method, which is an approximated, practical method, but
enables us to estimate the covariance matrix from the real data
itself. Another way is the method using mock catalogs of the universe
(lensing objects and weak lensing fields). In the following we define
the different covariance estimators considered in this paper, and then
discuss pros and cons of each method.

\subsubsection{``Full'' covariance based on the mock catalogs}
\label{sec:fullcov}

This should be a most accurate method if we have accurate mock catalogs
of the lensing observables (source galaxies and lensing galaxies or
clusters) for a given survey as well as a sufficient number of the
realizations for an assumed cosmological model.  Building such mock
catalogs requires adequately-designed numerical simulations.  However,
this is tricky and involves conflicting requirements. If a survey area
is wide, one has to include very large-scale modes in the light cone,
which requires a very large-box simulation. On the other hand,
simulating lens objects, either galaxies or clusters, requires an
adequately high-resolution simulation in order to resolve halos and/or
subhalos and then relate the halos/subhalos to lensing galaxies or
clusters.  In addition, an accurate estimation of the covariance matrix
requires a sufficiently large number of independent realizations of the
mock catalogs, further ideally as a function of cosmological
models. These requirements are not readily met by currently-available
computational resources.

Once the adequately accurate mock catalogs and their realizations are
available, we can estimate the covariance matrix using realizations of
the mock catalogs:
\beqa
 {\bf C}^{\rm full}(R_i,R_j)=\frac{1}{N_{\rm r}-1}\sum_{a=1}^{N_{\rm r}}
  \left[
\widehat{\Delta\Sigma}_{(a)}(R_i)-\overline{\Delta\Sigma}(R_i)
  \right]
  \left[
\widehat{\Delta\Sigma}_{(a)}(R_j)-\overline{\Delta\Sigma}(R_j)
	 \right]
  \label{eq:cov_full}
\eeqa
where $\widehat{\Delta\Sigma}_{(a)}(R_i)$ is the lensing profile at the
radial bin $R_i$ estimated from the $a$-th realization of the mock
catalog, and $N_{\rm r}$ is the number of the realizations.
$\overline{\Delta\Sigma}$ is the average of the lensing profiles over
all the realizations:
\beqa
 \overline{\Delta \Sigma}(R_i)=\frac{1}{N_{\rm r}}\sum_{a=1}^{N_{\rm r}}
  \Delta\Sigma_{(a)}(R_i).
\eeqa
We hereafter call the covariance estimator (equation~\ref{eq:cov_full}) ``full
covariance''. An advantage of this method is we can properly take into
account the SSC contribution, if the mock catalogs include the
super-survey modes for a given survey geometry. A disadvantage is that
the accuracy of the covariance estimation depends on the
accuracy of mock catalogs, which requires adequate
computational resources to perform suitable $N$-body simulations as well
as ray-tracing simulations.

Cosmic sample variances in measurements of the stacked lensing profile
$\widehat{\Delta \Sigma}$ arise from statistical properties of
large-scale structure in the light cone from an observer to source
galaxies. The difference from the cosmic shear statistics is that, in
addition to the matter distribution in the light cone, we need to model
sample variance arising from the distribution of lensing objects (halos
we will consider) and the cross-correlation with their surrounding
matter distribution at particular lens redshifts. In order to properly
model the statistical properties of stacked lensing, we will use a suite
of high-resolution $N$-body simulations as well as ray-tracing simulations
that are computed by tracing light rays through the simulated matter
distribution in the light cone, as we describe below.

\subsubsection{Jackknife method}
\label{sec:jackknife}

The jackknife (JK) method is one of the techniques most conventionally
used for estimating the covariance matrix in the literature.  The JK
method allows one to estimate the covariance matrix of stacked lensing
from the data itself. The implementation is as follows.  (1) Divide
the survey region into different $N_{\rm sub}$ subregions, where each
subregion is defined so as to have an equal (or roughly equal) area. (2)
Estimate the stacked lensing profile from the survey region, but
excluding each of the $N_{\rm sub}$ subregions.  We call this
measurement excluding the $\alpha$-th subregion ``the $\alpha$-th JK
resample''. (3) Repeat the (2) measurements for all the $N_{\rm sub}$
subregions, and then construct the $N_{\rm sub}$ JK resamples. By
construction, each JK resample consists of ($N_{\rm sub}-1$) subregions
that has an smaller area than that of the original dataset by a factor
of $(N_{\rm sub}-1)/N_{\rm sub}$. (4) The covariance matrix is estimated
from the $N_{\rm sub}$ JK resamples as
\beqa
 {\bf C}^{\rm JK}(R_i,R_j)=\frac{N_{\rm sub}-1}{N_{\rm sub}}
  \sum_{\alpha=1}^{N_{\rm sub}}\left[\widehat{\Delta\Sigma}_{(\alpha)}(R_i)-\overline{\Delta\Sigma}(R_i)\right]
  \left[\widehat{\Delta\Sigma}_{(\alpha)}(R_j)-\overline{\Delta\Sigma}(R_j)\right], 
\eeqa
where $\widehat{\Delta\Sigma}_{(\alpha)}(R_j)$ is the lensing
measurement for the $\alpha$-th JK resample, and
\beqa
 \overline{\Delta\Sigma}(R_i)\equiv
  \frac{1}{N_{\rm
  sub}}\sum_{\alpha=1}^{N_{\rm sub}}\widehat{\Delta\Sigma}_{(\alpha)}(R_j)
\eeqa
The concept of the JK method is the different JK resamples give
different realizations of the measurements, which then enable one to
estimate the covariance. It is known that the JK method gives an
unbiased estimator of the underlying true covariance if the mass and
galaxy distributions in large-scale structure obey a random Gaussian or
Poisson distribution. However, the fields are
non-Gaussian, and all the different modes are correlated with each
other. In particular, it has not been fully addressed whether the JK
method captures the SSC contribution for a given survey
region. Addressing this question is one of our main scopes.

The advantage of the JK method is that, since the JK covariance is
estimated from the actual dataset itself, it automatically incorporate
observational effects on the covariance estimation, e.g., possible
inhomogeneities in the depth of data and selection function of lensing
galaxies and/or source galaxies.  However, there are several caveats
that must be kept in mind.  First, the accuracy of JK covariance
estimation depends on the number of JK subregions. If an insufficient
number of JK subregions are taken, the covariance estimation becomes
noisy because the insufficient JK resamples cannot well capture the
underlying variations in the stacked lensing. Second, the size of JK
subregion sets a maximum separation scale of the stacked lensing profile
used for the cosmological analysis, because the JK resamples would by
construction display less variations in the stacked lensing profile at
larger separations than the JK subregion size. Since the stacked lensing
profile at large separations contains a cleaner cosmological information
in the weakly nonlinear or linear regime, we want to have an access to
such large separations.  Hence there is a trade-off in the JK
subdivisions of dataset: the number of JK subregions vs. the maximum
separation scale. If we want to use the large separation, we need to
take a relatively large-area JK subregion, yielding a relatively small
number of JK resamples, and then the covariance estimation becomes
noisy.

\subsubsection{``Subsample'' covariance method: the covariance scaled from
   subregion area to total area}
\label{sec:subcov}

In order to develop our understanding of the nature of JK covariance, we
also study another method, which we hereafter call the ``subsample''
covariance. We implement this method as follows. (1) Divide the survey
region into $N_{\rm sub}$ subregions, where each subregion has an equal
area as in the JK method. (2) Use different realizations of the
light-cone simulations (therefore independent large-scale structures) to
simulate lensing effects on source galaxies in each subregion. (3)
Measure the stacked lensing profiles from each subregion, and estimate
the covariance matrix from the standard deviations among the $N_{\rm
sub}$ measurements. (4) Estimate the covariance matrix for the total
area, by scaling the covariance matrix of subregions by a factor of the
area ratio as
\begin{eqnarray}
 {\bf C}^{\rm sub}_{ij}&\equiv &\frac{\Omega_{\rm sub}}{\Omega_{\rm s}}
\frac{1}{N_{\rm sub}-1}
  \sum_{\beta=1}^{N_{\rm sub}}\left[\widehat{\Delta\Sigma}_{(\beta)}(R_i)-\overline{\Delta\Sigma}(R_i)\right]
  \left[\widehat{\Delta\Sigma}_{(\beta)}(R_j)-\overline{\Delta\Sigma}(R_j)\right],
  \label{eq:cov_sub}
\end{eqnarray}
where $\Omega_{\rm s}$ is the area of the survey (total area),
$\Omega_{\rm sub}$ is that of each subregion ($\Omega_{\rm
s}/\Omega_{\rm sub}=N_{\rm sub}$ in our setting), and
$\widehat{\Delta\Sigma}_{(\beta)}(R_i)$ is the lensing profile measured
from the $\beta$-th subregion, and
\beqa
 \overline{\Delta\Sigma}(R_i)\equiv
  \frac{1}{N_{\rm
  sub}}\sum_{\beta=1}^{N_{\rm sub}}\widehat{\Delta\Sigma}_{(\beta)}(R_i).
\eeqa
The difference of this method from the JK method is that the lensing
profiles measured from the different subregions are independent, while
the lensing fields in the JK method arise from the same realization of
large scale structure covering the total area.

Besides the prefactor of $\Omega_{\rm sub}/\Omega_{\rm s}$ in
equation~(\ref{eq:cov_sub}), the estimator is similar to
equation~(\ref{eq:cov_full}), thus estimating the full covariance for the
subregion area. If the covariance amplitude scales with a survey area as
$1/\Omega_{\rm s}$, the covariance for the total survey area would be
smaller than that of the subregion by a factor of $\Omega_{\rm
sub}/\Omega_{\rm s}$. However, as we discussed, this scaling relation
should be violated in the presence of SSC. The SSC contribution to the
full covariance for subregion area arises from super-survey modes whose
length scales are greater or comparable with the size of ``subregion'',
not total survey area.  Hence by comparing the above covariance with the
full covariance method in Section~\ref{sec:fullcov}, we can quantify the
relative contribution of SSC terms of the subregion and total areas. We
expect the greater SSC contribution for the subregion area than that for
the total area, because the variance of the super-survey modes for the
subregion area is larger than that for the total area as predicted in
the CDM scenario \citep{2013PhRvD..87l3504T}. Note that, as in the JK
method, a maximum separation $R_{\rm max}$ of the lensing profile
$\Delta\Sigma(R)$ is set by the size of the subregion. This subsample
covariance would be the case if different subregions are separated by a
large distance on the sky such that the subregions are considered
independent (no correlation between the lensing fields in different
subregions)\footnote{Exactly speaking, even in this case the different
subregions are correlated by super-survey modes corresponding to the
separation distance, so this causes the SSC term, which should however
be smaller than the SSC term of subregion area.}.

\change{
By the analogy to clustering analysis of galaxies, 
one may be concerned about the integral constraint in the case of 
subsample covariance method.
Nevertheless, the stacked shear profile, which is defined in terms of the integral of the power spectrum weighted with the 2nd-order Bessel function, $J_2(x)$ (see Eq.~[\ref{eq:dsigma_theory}]),
does not obey the integral constraint, unlike the two-point correlation function of the convergence field that involves the integral of the zero-th order Bessel function, $J_0(x)$ 
(e.g., see around Eq.~(1) in \citet{2011PhRvD..83j3509B}). 
Hence, the stacked shear profile even for the subregion method can be estimated without any problem, based on the pair counting statistics.
}

\section{SIMULATIONS}\label{sec:sim}
\subsection{$N$-body simulation}

In order to study in detail the weak lensing statistics considered in
this paper, we use the full-sky weak lensing maps that are constructed
from multiple sets of $N$-body simulation realizations.  $N$-body
simulations can properly model nonlinear structure formation as well as
the resulting non-Gaussianity in the distributions of halos and dark
matter.

To run each cosmological $N$-body simulation, we used the parallel
Tree-Particle Mesh code {\tt Gadget2} \citep{Springel2005}.  We employed
$2048^3$ $N$-body particles.  For each $N$-body simulation, we generated the
initial conditions using a parallel code developed by
\citet{2009PASJ...61..321N} \citep[also see][]{2011A&A...527A..87V},
which is based on the second-order Lagrangian perturbation theory
\cite[e.g.][]{2006MNRAS.373..369C}.  To set up the the initial
conditions we used {\tt CAMB} \citep{Lewis2000} to compute the linear
matter transfer function for the fiducial cosmological model.  The
fiducial model is characterized by the following cosmological
parameters: the density parameter of total matter $\Omega_{\rm
m0}=0.279$, the density parameter of baryon $\Omega_{\rm b0}=0.046$, the
density parameter of the cosmological constant $\Omega_{\Lambda
0}=0.721$, the density fluctuation amplitude $\sigma_{8}=0.823$, the
Hubble parameter $h=0.700$ and the spectral index $n_s=0.972$.  These
parameters are consistent with the nine-year WMAP results
\citep{Hinshaw2013}.

To simulate large-scale structures in the light cone covering up to
sufficiently high redshifts of source galaxies we are interested in, we
used the $N$-body simulation realizations for 9 different box sizes
which range from $450\, h^{-1}$Mpc to $4050\, h^{-1}$Mpc, stepped by
$450\, h^{-1}$Mpc in the side length.  For each box-size $N$-body
simulation, we generated 6 different realizations, each of which was run
using the different initial seeds for the same cosmological model. Hence
we use $54(=6\times 9)$ $N$-body simulation realizations in total.  As
illustrated in Fig.~2 of \citet{2015MNRAS.453.3043S}, we place the
$N$-body simulation realizations around a hypothetical observer at the
coordinate origin (``$O$'' in the left panel of
Fig.~\ref{fig:light_cone}) so as to cover the past light-cone around the
observer.  To cover the full sky, we repeatedly use the same realization
of each box size, by placing 8 boxes of the same realization around the
observer. In doing this the matter distribution in each redshift slice
becomes continuous due to the periodic boundary condition.  However,
note that the ray-tracing simulations we use ignore the Fourier modes
greater than the $N$-body simulation volume, which correspond to angular
modes greater than $\sqrt{4\pi/8}$ radian.  The largest-volume
simulation with $4050\, h^{-1}$Mpc on a side enables us to simulate the
lensing effect on source galaxies up to redshift $z\simeq 2.4$ for our
fiducial cosmological model.  While one main purpose of this paper is to
generate a mock catalog of the SDSS weak lensing fields, the redshift
distribution of SDSS source galaxies peaks around $z\simeq 0.5$ and has
a long tail extending up to $z\sim 1$, as shown in Fig.~5 in
\citet{2013MNRAS.432.1544M} (more exactly speaking the maximum
photometric redshift is $z\simeq 1.49$). Thus our $N$-body simulations
are sufficiently large to cover the entire large-scale structure in the
light cone that is probed by SDSS (also see blow).  The further details
can be found in \citet{2015MNRAS.453.3043S} (also see Takahashi et
al. in preparation).

\subsection{Full-sky lensing simulation}\label{subsec:fullsky_lens}

We here briefly summarize our ray-tracing simulations with full-sky
coverage.  The detailed description is found in Appendix~A of
\citet{2015MNRAS.453.3043S}.  In our ray-tracing simulation, we used the
standard multiple lens-plane algorithm \citep[e.g.,][]{Hamana2001} to
compute the light-ray path and the lensing magnification matrix for
source galaxies in different redshift slices.  To take into account the
curvature of the celestial sphere, we performed the multiple lens-plane
algorithm on a spherical geometry using the projected matter density
field given in the format of spherical shell, which is generated from
$N$-body simulation realization output at the redshift corresponding to
the distance to the shell from an observer.  We used the projected
matter fields in 27 shells in total, each of which was computed by
projecting $N$-body simulation realization over a radial width of $150\,
h^{-1}$Mpc, in order to make the light cone covering a cosmological
volume up to $z\simeq 2.4$.  As described above, we used 6 different
box-size $N$-body simulations.  For each box-size simulation, we
computed the projected matter fields in 3 shells of different radii
(radial distances), where the projected matter field in each shell is
computed from different regions of the $N$-body simulation output at the
corresponding redshift and the largest-radius shell is inscribed to the
simulation box \citep[see Fig.~2 of][]{2015MNRAS.453.3043S}.  Hence, our
lensing simulations can model the evolution of the density field along
the line of sight with the interval of comoving $150\,h^{-1}$Mpc,
corresponding to the redshift width of $\sim0.05-0.1$.
Table~\ref{tab:each_density_shell} summarizes the redshifts of $N$-body
simulation outputs.

\begin{table*}
\begin{tabular}{@{}llcccc|}
\hline
\hline
$L_{\rm box}\, [h^{-1}{\rm Mpc}]$  & ID of lens-shells & Output redshifts & Number of realizations \\ \hline
 450
 & 1, 2, 3  & 0.0251, 0.0763, 0.128 & 6 \\
 900 
 & 4, 5, 6  & 0.182, 0.237, 0.294 & 6 \\
 1350 
 & 7, 8, 9  & 0.352, 0.413, 0.475 & 6 \\
 1800 
 & 10, 11, 12  & 0.540, 0.607, 0.677 & 6 \\
 2250 
 & 13, 14, 15  & 0.750, 0.827, 0.906 & 6 \\
 2700 
 & 16, 17, 18  & 0.990, 1.077, 1.169 & 6 \\ 
 3150 
 & 19, 20, 21  & 1.267, 1.369, 1.477 & 6 \\ 
 3600 
 & 22, 23, 24  & 1.592, 1.714, 1.844 & 6 \\ 
 4050 
 & 25, 26, 27  & 1.982, 2.129, 2.287 & 6 \\ 
\hline
\end{tabular}
 \caption{Parameters of $N$-body simulations, used in generating the
 full-sky light-cone simulations.  Each simulation was run with $2048^3$
 dark matter particles.  The output redshift of each simulation
 corresponds to the comoving distance to the center of lens shell. We
 adopted the standard CDM model, which is consistent with WMAP 9-yr
 results \citep{Hinshaw2013}. } \label{tab:each_density_shell}
\end{table*}

\begin{figure*}
\centering
\includegraphics[width=0.4\columnwidth, bb = 0 0 366 348]
{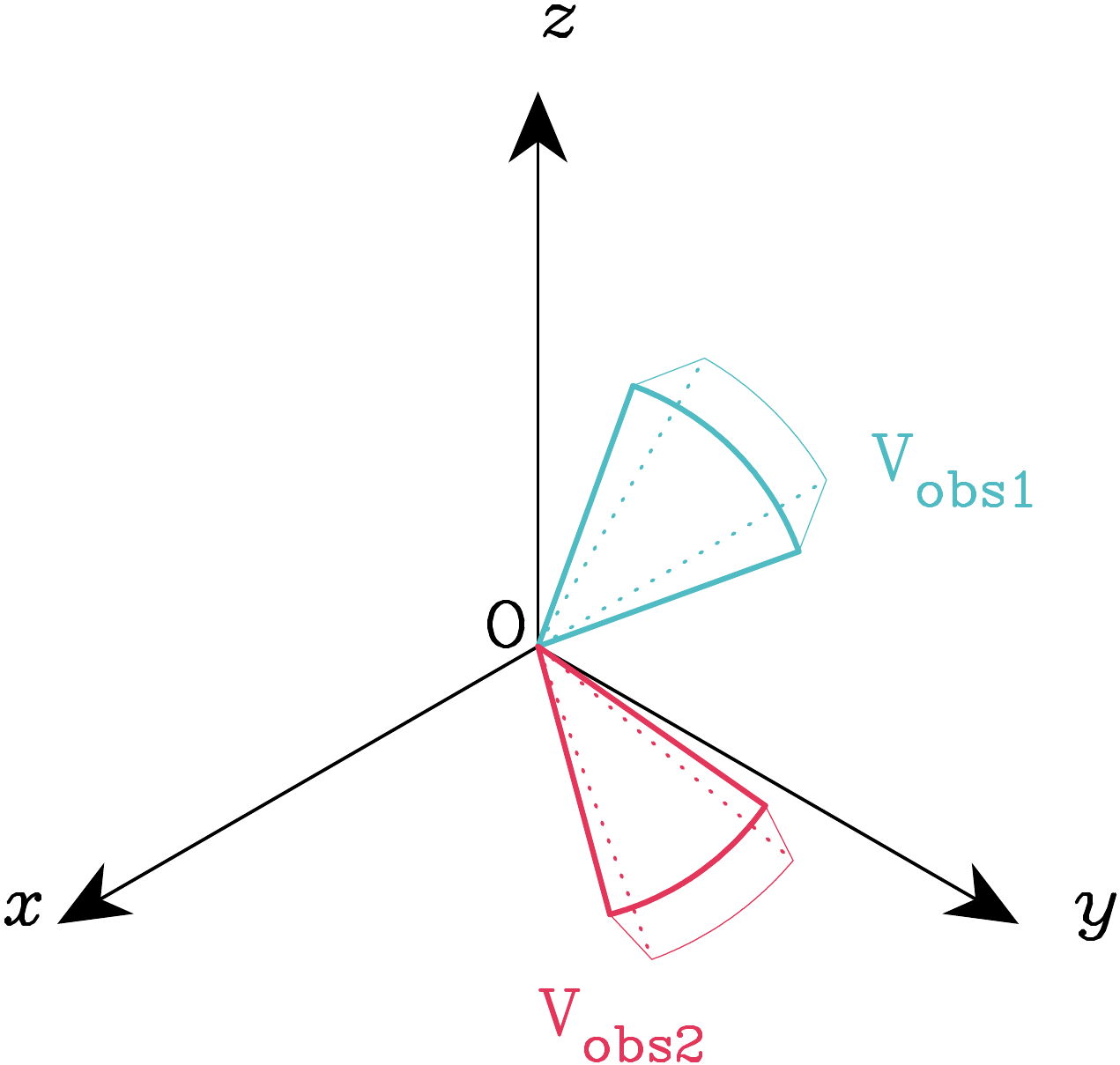}
\includegraphics[width=0.47\columnwidth, bb = 0 0 523 404]
{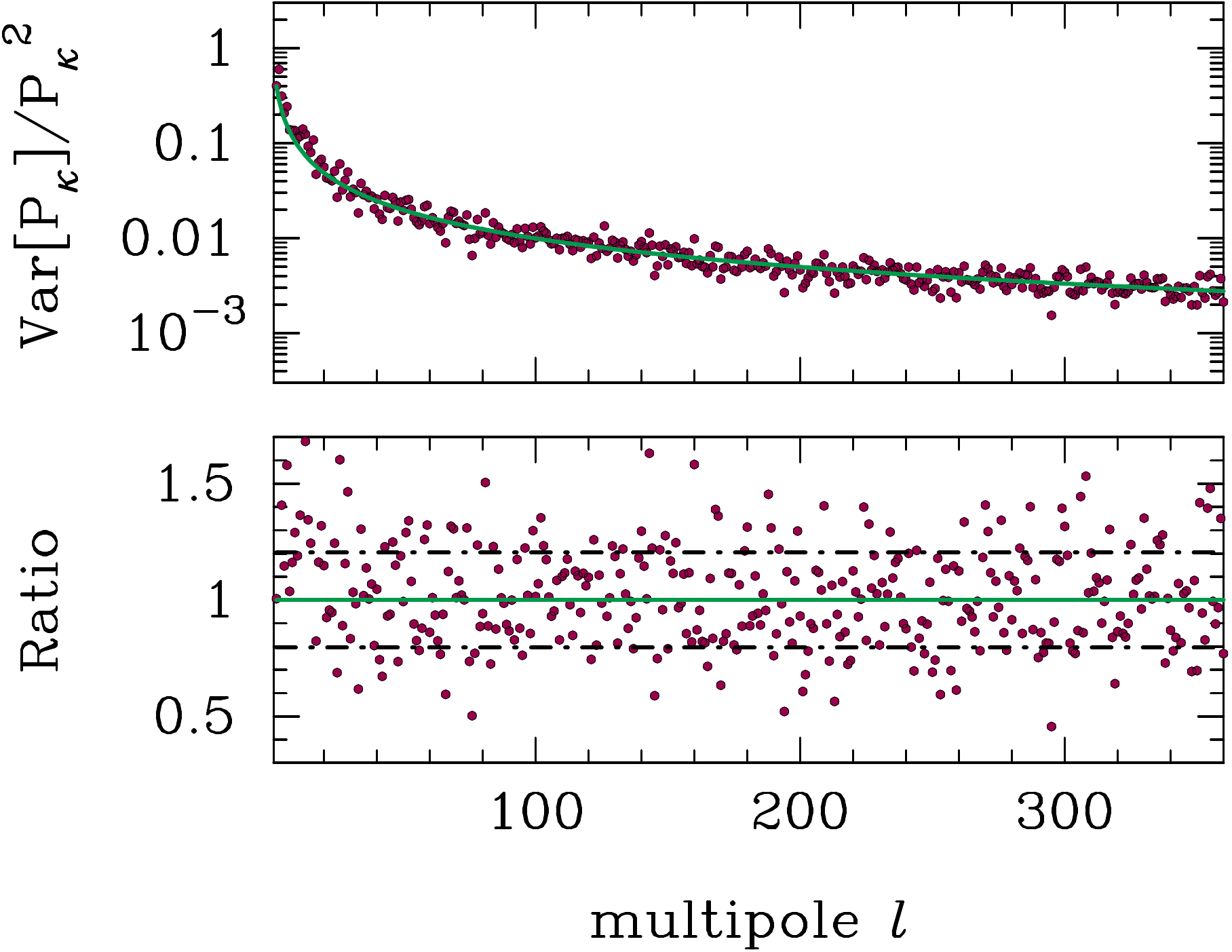}
\caption{{\it Left figure}: An illustration of a galaxy survey with
partial sky coverage, which has a volume of $V_{\rm obs}$, in the
light-cone simulation. A hypothetical observer is at the coordinate
origin, $O$. {\it Right panel}: The upper plot shows the power spectrum
variance for the cosmic shear convergence field,
estimated from the spectra measured from the 48
full-sky lens maps.
Here we consider the source redshift $z_s=1$ and the range of low
multipoles, $\ell\simlt360$, in the linear regime.  We show the
variance normalized by the average power spectrum squared, compared to
the Gaussian variance $2/(2\ell+1)$ (green curve).  In the bottom panel,
we show the ratio of the measured variance and the Gaussian prediction
and the black dashed lines shows the expected Gaussian error of the
variance, i.e., $\sqrt{2/N_{\rm rea}}$ where $N_{\rm rea}=48$.
}
\label{fig:light_cone}
\end{figure*} 
In computing the projected matter fields from each $N$-body realization,
we randomly shift and rotate the simulation box in order to avoid the
same structure appearing multiple times along the same line of
sight. Here we properly computed the deflection of light ray in each
lens plane; we did not employ the Born approximation.  We used the {\tt
HEALPix} software \citep{2005ApJ...622..759G} to generate the projected
matter fields in pixels embedded into each of the spherical shells.  The
angular resolution parameter $nside$ is set to be 8192 corresponding to
$\simeq0.43$~arcmin for the pixel scale (angular
resolution)\footnote{For a given $nside$, the {\tt HEALPix} software
divides the full sky into $12\times nside^2$ pixels each of which has an
equal area, and therefore the area of each pixel is given by
$4\pi/(12\times nside^2)$.}.  In our simulation, we simulated light ray
trajectories from $z=0$ to $z\simeq 2.4$ using the standard multiple
lens-plane algorithm.  To efficiently generate mock catalogs, we stored
the data of weak lensing fields (convergence and shear) in each of 26
shells for each full-sky realization.  Thus the source planes are
located in radial distances ranging from $150\, h^{-1}$Mpc to $4050\,
h^{-1}$Mpc, stepped by $150\, h^{-1}$Mpc, while the lens planes are
placed at a cone-volume weighted mean radial distance of each shell.
In this paper, we use the weak lensing fields in
21 shells corresponding to the maximum redshift of SDSS source galaxies,
$z_{\rm s}\simeq 1.49$. In the following we use 48 realizations of the
full-sky lensing maps that are constructed from different combinations
of 54 $N$-body realizations (6 realizations for each of 9 different-size
$N$-body simulations).

Our full-sky lens maps are built by using the same set of $N$-body
simulation realizations multiple times. However, when making mock
catalogs of a ``partial-sky-coverage'' lens map, each realization of the
lensing maps is taken from the partial-solid-angle light cone around a
hypothetical observer, corresponding to different subregions of $N$-body
simulation realizations, as illustrated in Fig.~\ref{fig:light_cone}.
The mock catalogs generated from subregions of each $N$-body realization
allow us to study the effects in galaxy-galaxy weak lensing for a
finite-volume survey that is embedded in a much larger volume universe
\citep[see][for the similar study for three-dimensional matter power
spectrum]{Lietal:14a}.  First, the large-scale structure ($N$-body
particle distribution) simulated in each subregion of $N$-body
simulation realization arises from independent, linear initial density
fields with length scales smaller than the subregion scale. Secondly,
the matter clustering in each subregion includes mode-coupling effects
with super-survey modes beyond the subregion size.  As long as a volume
of subregion is large enough, the super-survey modes are in the linear
regime, and therefore the different subregions are affected by
independent super-survey modes. Thus our simulated mock catalog mimics
what we observe for an actual survey.  If mock catalogs are generated
from $N$-body simulations with periodic boundary conditions
(i.e. without super-survey modes), the mock catalogs do not include the
effects of super-survey modes. 

We consider that the different realizations of our full-sky lens
maps are nearly independent for the following reason.  As we stated, we
generate the different full-sky lens maps based on the multiple lens
plane algorithm, where we project the three-dimensional $N$-body
particle distribution to each lens shell along the line-of-sight
direction, after performing random translation and rotation of each
$N$-body realization. Since the lensing fields arise from Fourier modes
perpendicular to the line-of-sight direction, the random translation and
rotation effectively allow us to make independent realizations
\citep{Hamana2001}. Thus in this was we can construct many, nearly
independent realizations of the full-sky lens maps from different
combinations of of 54 different-box $N$-body simulation realizations.
As a justification of independence of the full-sky maps, in the right
panel of Fig.~\ref{fig:light_cone} we examine the cosmic shear power
spectrum, $P_\kappa(\ell)$, for source redshift at $z_s=1$ and in the
multipole range of $\ell\simlt360$ from each full-sky realization, where
the low multipole range is considered to be in the linear regime.  Then
we estimate its covariance matrix from the 48 full-sky realizations.
The figure shows that the variance of $P_{\kappa}$ is consistent with
the Gaussian prediction of ${\rm Var}(P_{\kappa})=2P_{\kappa}^2/(2\ell
+1)$ (we took $\Delta \ell=1$ in this calculation).  Furthermore, the
scatters in the variance are also found to be close to the Gaussian
prediction of $\sqrt{2/N_{\rm rea}}{\rm Var}(P_{\kappa})$ where $N_{\rm
rea}=48$ \citep{Tayloretal:13}.  These results support that our full-sky lens maps can be
regarded as quasi-independent realizations. Furthermore, when
considering a lensing survey with a partial sky coverage, we can
construct the different realizations by clipping a finite solid-angle
light cone from the full-sky map along different line-of-sight
directions from the observer (see the left illustration of
Fig.~\ref{fig:light_cone}), where the different realizations are from
different subvolumes of the original $N$-body simulations in the light
cone. Thus, we can further increase the number of independent
realizations.

\subsection{Full-sky halo catalog}
\label{subsec:fullsky_halo}

In order to simulate the stacked lensing, i.e. galaxy- or cluster-shear
cross-correlations, we also need a mock catalog of lensing halos that
host galaxies and/or clusters we can observe from real data. We
implemented the software {\tt Rockstar} \citep{2013ApJ...762..109B} to
each output of $N$-body simulation in the post processing, where {\tt
Rockstar} identifies dark matter halos from a clustering of dark matter
particles in phase space.  Throughout this paper, we define the halo
mass by using the spherical overdensity criterion: $M_{\rm
200m}=200\bar{\rho}_{\rm m0}(4\pi/3)R^{3}_{\rm 200m}$, where
$\bar{\rho}_{\rm m0}$, the present-day mean mass density, is due to our
use of the comoving coordinates.  For redshifts $z_{\rm l}\simlt 0.5$
relevant for lensing galaxies/clusters, our $N$-body simulations allow
us to resolve dark matter halos with masses greater than a few times
$10^{12}h^{-1}\, M_\odot$ with more than 50 $N$-body particles.  In the
following we use only ``host'' halos and do not use subhalos. We believe
that this does not affect our results, because the main goal of this
paper is to study the nature of the covariance matrix of galaxy-galaxy
weak lensing in the sample variance limited regime that is in the
two-halo term regime.  We again use the {\tt HEALPix} to assign the
position of each dark matter halo to the pixels in the celestial sphere.
Although we used the halo catalogs at output redshift corresponding to
each shell of the projected density fields (150, 300,
... 4050~$h^{-1}{\rm Mpc}$ stepped by 150~$h^{-1}{\rm Mpc}$ as given in
Table~\ref{tab:each_density_shell}), we used the radial distance of each
halo from an observer to assign the corresponding redshift to the halo,
assuming the fiducial cosmological model. However, this redshift
information is a small effect, and we checked that the following results
are not changed even if we use the redshift of lens plane for all halos
residing in the same lens slice.

\subsection{Generating the mock catalogs of galaxy-galaxy weak lensing}
\label{subsec:mock}

\begin{figure*}
\centering
\includegraphics[width=0.32\columnwidth, bb=0 0 445 336]
{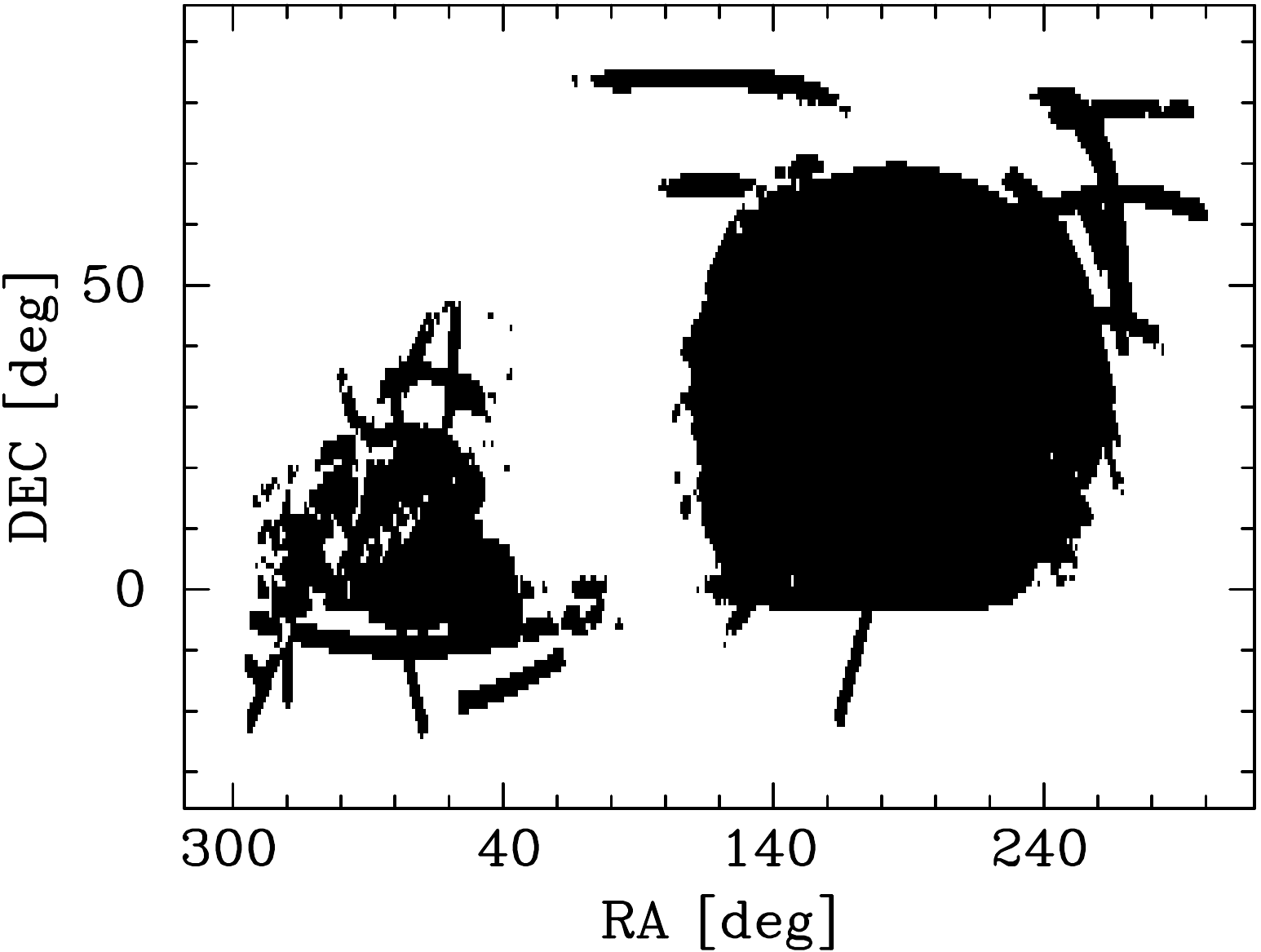}
\includegraphics[width=0.32\columnwidth, bb=0 0 445 336]
{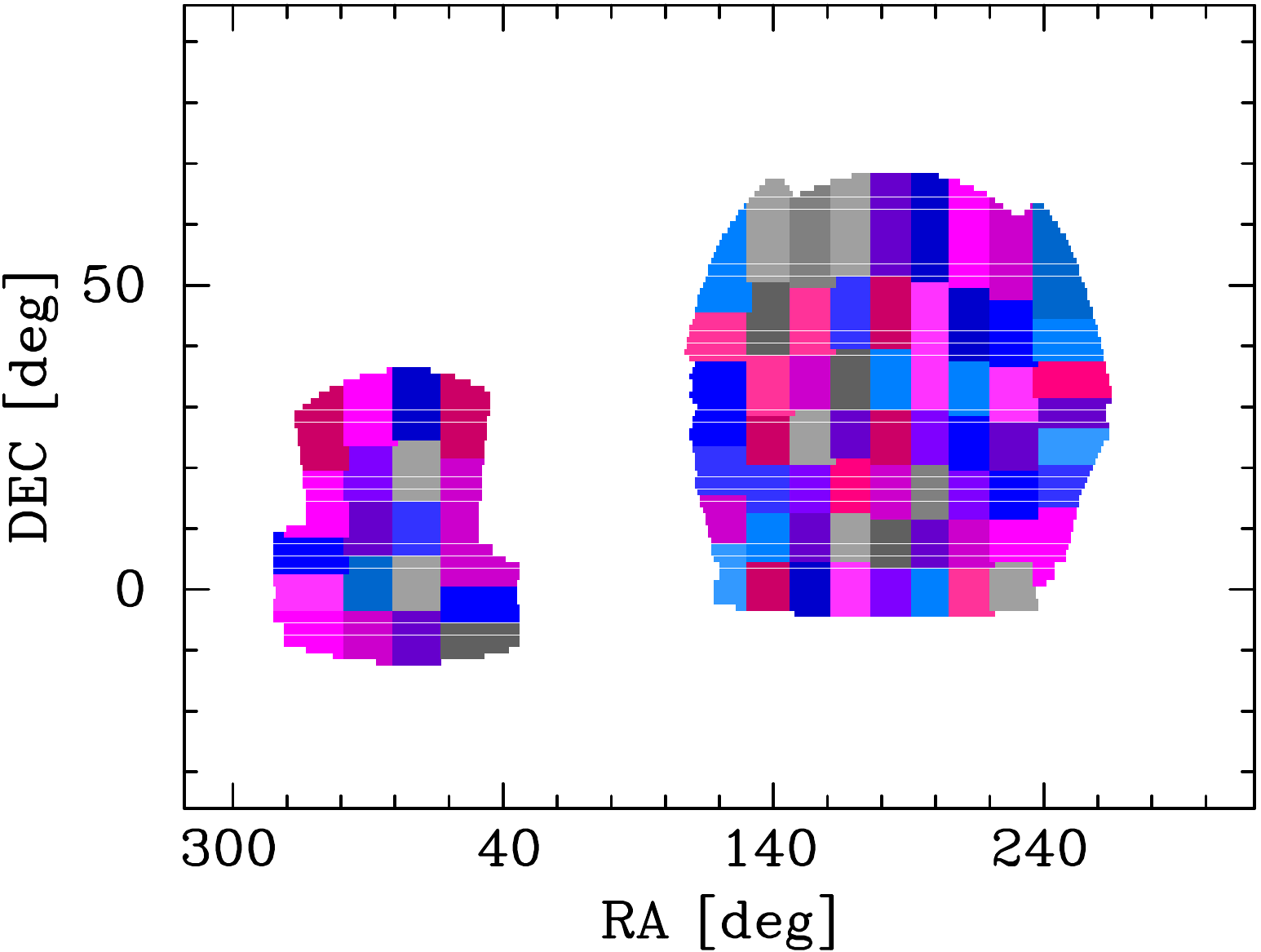}
\includegraphics[width=0.32\columnwidth, bb=0 0 445 336]
{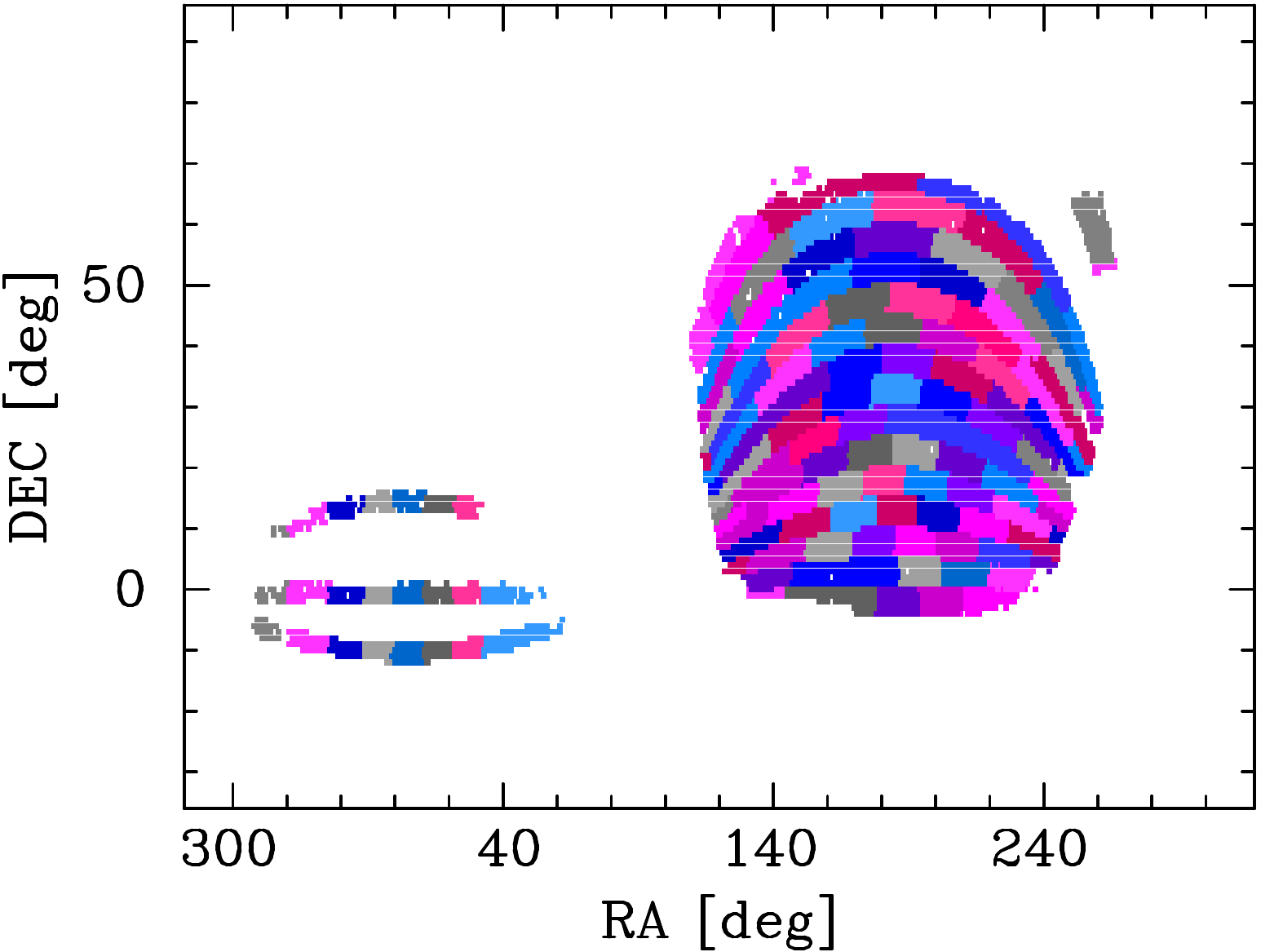}
 \caption{The survey footprints of the SDSS data used in this paper.
 {\it Left panel}: The footprints of the source galaxies constructed
 from the SDSS imaging data \citep{2012MNRAS.425.2610R}, which we use
 for the weak lensing measurements.  {\it Middle}: The footprints of
 redMaPPer clusters \citep{2014ApJ...785..104R, 2015MNRAS.453...38R},
 which we use as tracers of large-scale structure for the stacked
 lensing measurements.
 The different color regions denote 83 roughly-equal-area subregions
 used for the jackknife covariance estimation, following the method in
 \citet{2016PhRvL.116d1301M}.  {\it Right}: The footprints of luminous
 red galaxies (LRGs) which we use as another large-scale structure
 tracers for the galaxy-galaxy weak lensing measurements. The different
 color regions denote 100 subregions for the jackknife covariance
 estimation following \citet{2013MNRAS.432.1544M}.  }
 \label{fig:sdss_footprint}
\end{figure*} 

In this section, we describe the details of creating mock catalogs for
galaxy- or cluster-shear cross-correlations with full-sky simulations in
Section~\ref{subsec:fullsky_lens} and \ref{subsec:fullsky_halo}.  Our
primary focus is to demonstrate how we can use the full-sky simulations
to perform mock observations of stacked lensing in the
currently-available largest sky surveys, the Sloan Digital Sky Survey
\citep[SDSS;][]{2000AJ....120.1579Y}, which covers about one-quarter of
the sky.  Our mock catalogs are generated by combining the real catalog
of source galaxies in the SDSS, the full-sky ray-tracing simulations and
the halo catalogs. The computation procedures are summarized as follows:
\begin{itemize}
 \item Take the real catalog of source galaxies, where each galaxy
       contains information on the position (RA and dec), shape,
       redshift and the lensing weight. Populate each source galaxy into
       one realization of our light-cone simulations according to its
       angular position and redshift.  Thus our method maintains the
       observed distribution of source galaxies on the sky, their shapes
       as well as their redshift distribution.
 \item Randomly rotate ellipticity of each source galaxy to erase the
       real lensing signal.
 \item Simulate the lensing distortion effect on each source galaxy by
       adding the lensing contribution at each foreground lens plane,
       weighted by the lensing efficiency $\Sigma_{\rm cr}(z_{\rm
       l},z_{\rm s})$, to the intrinsic shape. 
 \item Identify halos that are considered to host galaxy clusters or
       galaxies, the redMaPPer clusters or the Luminous Red Galaxies
       (LRGs) in this paper, in the light-cone coordinates (redshift and
       angular position of each halo).
 \item Make hypothetical measurements of cluster- or galaxy-shear
       cross-correlations following exactly the similar method used in
       the actual measurements.
\end{itemize}
In the following we describe details of the SDSS shape catalog and the
mock catalogs of the redMaPPer clusters and LRGs.

\subsubsection{SDSS shape catalog}
\label{subsec:sdss_shape}

To model source galaxies, we use the shape catalog that was constructed
in \citet{2012MNRAS.425.2610R} from the SDSS DR8 photometric data
\citep{Aiharaetal:11}.  In this catalog, the galaxy shapes are measured
by the re-Gaussianization technique \citep{2003MNRAS.343..459H}, and the
systematic uncertainties involved in the shape measurements have been
investigated in detail \citep{2005MNRAS.361.1287M}. In this catalog the
inverse variance weight, $w_{\rm s}$, is assigned to each source galaxy
as
\beqa
w_{\rm s} = \frac{1}{\sigma^{2}_{\epsilon}+\sigma^2_{\rm SN}},
\eeqa
where $\sigma_{\epsilon}$ is the estimated shape measurement error and
$\sigma_{\rm SN}$ represents the rms intrinsic ellipticity, for which we
throughout this paper employ $\sigma_{\rm SN}=0.365$ according to the
result in \citet{2012MNRAS.425.2610R}.  Note that, when computing the
stacked lensing, we further include the lensing efficiency for the
weight, $\Sigma_{\rm cr}(z_{\rm l},z_{\rm s})$ for each source-lens
pair, as given in equation~(\ref{eq:w_ls}).  
Following \citet{2013MNRAS.432.1544M}, we use the maximum-likelihood mode of {\tt
ZEBRA} and choose the best-fitting photo-$z$, after marginalizing over
the SED template, for each source galaxy redshift.  As described in
detail in \citet{2012MNRAS.425.2610R} \citep{2013MNRAS.432.1544M}, we
use 39,267,029 source galaxies in total, that remain after the
carefully-tuned selection based on the imaging quality, data reduction
quality, galactic extinction, apparent magnitude, photometric redshift
and galaxy size.

To simulate weak lensing effects on the SDSS source galaxies, we use the
similar method developed in \citet{Shirasaki2014}. In doing this, we
keep all the characteristics of the SDSS data; the angular positions,
the redshift and ellipticity distributions, and the survey geometry.

We first assign the celestial coordinates (RA and dec) to the full-sky
simulation map. Then we populate the SDSS source galaxies in the
full-sky map as follows. (1) We assign each source galaxy to the
redshift of the nearest source plane from its photometric redshift, and
then assign the galaxy to the nearest pixel in the source plane (the
pixel scale is about $0.43^{\prime}$).  Note that, as we described, our
source planes are discrete and located in the radial distances starting
from $150\,h^{-1}{\rm Mpc}$ up to $4050\, h^{-1}{\rm Mpc}$, stepped by
$150\, h^{-1}{\rm Mpc}$, corresponding to the redshift width of $\Delta
z\simeq0.05$. Since this redshift width is smaller than a typical error
of the photometric redshifts of SDSS galaxies (by a factor of 2)
\citep{2012MNRAS.420.3240N}, we believe that the assignment procedure of
source redshifts would be accurate enough. More exactly speaking we can
further include the probability distribution of photo-z for each galaxy
in the redshift assignment, but we ignore this for simplicity in this
paper.  (2) For each galaxy we randomly rotate the ellipticity
orientation to erase the real lensing signal.  Since the intrinsic
ellipticity is larger than the lensing effect on individual galaxy
basis, we can approximate the intrinsic ellipticity, ${\bm
\epsilon}_{\rm int}$, by the observed ellipticity.  We should note that
the ellipticity of the shape catalog in \citet{2012MNRAS.425.2610R} is
defined in terms of the {\it distortion}
\citep{1991ApJ...380....1M,BernsteinJarvis:02}, whose amplitude is
related to the major and minor axes of an ellipse as
$(a^2-b^2)/(a^2+b^2)$.  (3) We then simulate the lensing distortion
effect on each galaxy. As we described, we have the lensing fields as a
function of the angular position (RA and dec) and source plane redshift.
Using the matrix multiplication rule, we can simulate the ``observed''
ellipticity, denoted as ${\bm \epsilon}_{\rm mock}$, for each galaxy by
adding the intrinsic ellipticity and the simulated lensing effect:
\beqa \epsilon_{\rm mock,
1} &=& \frac{\epsilon_{\rm int, 1}+\delta_{1}+(\delta_{2}/\delta^2)
[1-(1-\delta^2)^{1/2}] (\delta_{1}\epsilon_{\rm int,
2}-\delta_{2}\epsilon_{\rm int, 1})} {1+{\bm \delta}\cdot{\bm
\epsilon}_{\rm int}}, \\ \epsilon_{\rm mock, 2} &=& \frac{\epsilon_{\rm
int, 2}+\delta_{2}+(\delta_{1}/\delta^2) [1-(1-\delta^2)^{1/2}]
(\delta_{2}\epsilon_{\rm int, 1}-\delta_{1}\epsilon_{\rm int, 2})}
{1+{\bm \delta}\cdot{\bm \epsilon}_{\rm int}},
\eeqa
where ${\bm \delta}\equiv 2(1-\kappa){\bm
\gamma}/[(1-\kappa)^2+\gamma^2]$ and $\kappa$ and $\gamma$ are the
lensing convergence and shear fields, respectively. Note ${\bm
\delta}\simeq 2{\bm \gamma}$ in the weak lensing regime.

Thus our mock source galaxies preserve the observed angular positions,
redshifts, intrinsic ellipticities and lensing weights of real galaxies,
while we properly simulate the ellipticity of galaxies by combining the
lensing distortion extracted from ray-tracing simulations with the
intrinsic ellipticity for individual galaxies. The intrinsic
ellipticities give a dominant source of the statistical errors in the
stacked lensing at small scales, so using the observed information of
source galaxies is crucial to reproduce the statistical errors at the
small scales as we
will show below.

\subsubsection{Clusters of galaxies -- redMaPPer clusters}
\label{subsec:redmapper}

The stacked lensing analysis requires to use tracers of large-scale
structure at lens redshift. As one kind of tracers, we use the publicly
available catalog of galaxy clusters, the red-sequence Matched-filter
Probabilistic Percolation (redMaPPer) catalog
\citep{2014ApJ...785..104R}, constructed based on the SDSS DR8
photometric galaxy catalog.  The redMaPPer algorithm identifies clusters
of galaxies as overdensities of red-sequence galaxies at similar
redshift by using the $ugriz$ magnitudes and their errors.  For each
cluster, the catalog contains an optical richness estimate $\lambda$, a
photometric redshift estimate $z_{\lambda}$, as well as the position and
probabilities of 5 candidate central galaxies.  In this paper, we
utilize the catalog created by v5.10 version of the algorithm with
various improvements \citep[see][for details]{2015MNRAS.453...38R}. This
is the same catalog as used in \citet{2016PhRvL.116d1301M}, which
consists of an approximately volume-limited sample of 8,648 redMaPPer
clusters with $20<\lambda<100$ and $0.1<z_\lambda<0.33$.

In order to create the mock catalog of redMaPPer clusters, we need to
model the relation between the observed optical richness and halo mass,
which allows us to assign halos in the simulations to hypothetical
redMaPPer clusters.  We adopt the relation derived in
\citet{2016arXiv160306953S}.  They assumed a log-normal distribution
$P(\ln M|\lambda)$, which describes the probability distribution of
cluster masses $\ln M$ for a given richness $\lambda$, and then
calibrated the mean relation $\langle M|\lambda \rangle$ using the weak
lensing measurements for the 4 subsamples divided by different ranges of
richness $\lambda$:
\beqa
\langle M|\lambda \rangle = 
10^{14.344} \, [h^{-1}{\rm M_{\odot}}] 
\left(\frac{\lambda}{40}\right)^{1.33}. \label{eq:mean_redMaPPer}
\eeqa
We also adopt the model variance in $\ln M$ as a function of richness, derived in
\citet{2016arXiv160306953S}:
\beqa
{\rm Var}(\ln M|\lambda) = \frac{1.33^2}{\lambda} 
+ \sigma^{2}_{\ln M|\lambda}, \label{eq:var_redMaPPer}
\eeqa
where the intrinsic scatter $\sigma_{\ln M|\lambda}$ is fixed to be 0.25
according to \citet{2014ApJ...783...80R} \citep[also
see][]{2015MNRAS.450..592R}. Convolving the observed number counts of
clusters with the above probability distribution between halo mass and
optical richness, we can infer the underlying halo mass function in the SDSS
area:
\beqa
 F^{\rm RM}(\ln M)=\frac{1}{N_{\rm tot}^{\rm RM}}
  \int_0^\infty\!\mathrm{d}\lambda~\frac{\mathrm{d}N^{\rm
  RM}}{\mathrm{d}\lambda}P(\ln M|\lambda),
\eeqa
where $(\mathrm{d}N^{\rm RM}/\mathrm{d}\lambda)~\mathrm{d}\lambda$ is
the observed number counts of redMaPPer clusters in the richness bin
$[\lambda,\lambda+\mathrm{d}\lambda]$, $N^{\rm RM}_{\rm tot}$ is the
total number of redMaPPer clusters (i.e. $N^{\rm RM}_{\rm tot}=8,648$),
and $F^{\rm RM}(\ln M)$ is a fraction of the underlying halo candidates
hosting redMaPPer clusters in halo mass bin $[M,M+\mathrm{d}\ln M]$
relative to the total number of redMaPPer clusters.

We employ the following approach to create a mock catalog of redMaPPer
clusters in each realization of our full-sky, light-cone
simulations. (1) We identify candidate halos in the simulation
realization that have $M_{\rm 200m}\ge 10^{13}~h^{-1}M_\odot$ and are in the
range of $0.1<z<0.33$, and then compute the fractional mass function
$F^{\rm sim}(\ln M)$, which is a fraction of halos in mass bin
$[M,M+\mathrm{d}\ln M]$ among all the selected halos. Since the range of
halo masses in the simulation is different from that in the redMaPPer
clusters, we renormalize the mass function $F^{\rm sim}(\ln M)$; we
multiply $F^{\rm sim}(\ln M)$ by an overall constant factor so as to
match the observed mass function, $F^{\rm RM}(\ln M)$, at high mass
ends, $M_{\rm 200m}\simgt $ a few $10^{14}~h^{-1}M_\odot$.  (2) If $F^{\rm
RM}(\ln M)\ge F^{\rm sim}(\ln M)$ in a halo mass bin [$M,M+\mathrm{d}\ln
M$], we select all the halos as hypothetical redMaPPer clusters. If
$F^{\rm RM}(\ln M)<F^{\rm th}(\ln M)$, we make a random downsampling of
halos in the mass bin.  (3) We also assign a hypothetical richness to
each selected halo assuming the probability $P(\lambda|\ln M)\propto
P(\ln M|\lambda)(\mathrm{d}N^{\rm RM}/\mathrm{d}\lambda)$ such that the
selected halos with assigned richness  reproduce the observed
richness function, $\mathrm{d}N^{\rm RM}/\mathrm{d}\lambda$.
(4) We then compute the ``normalized'' radial distribution of mock
halos, $(1/N^{\rm sim}_{\rm tot})\mathrm{d}N^{\rm sim} /\mathrm{d}\chi$,
in 20 discrete radial bins in the range of $0.1<z<0.33$ via the
redshift-distance relation $\chi(z)$ for the fiducial cosmological
model.  If the probability of mock clusters in a given radial bin is
greater than that of real redMaPPer clusters, we made a random
downsampling of mock clusters at the bin by the ratio.  For the
random catalogs of redMaPPer clusters, we used 20 realizations of the
random catalog that are used for the stacked lensing measurements from
the mock catalogs (see equation~\ref{eq:est_delta_Sigma}).

Next we define the survey footprints of SDSS data in the light-cone
simulation by using the randoms catalog of the redMaPPer clusters
(version 5.10 at the url \url{http://risa.stanford.edu/redmapper/}).  To
do this, we assign, to each full-sky simulation, the {\tt HEALPix}
pixels with angular resolution parameter $nside=512$ corresponding to
angular resolution of about 6.8~arcmin, which is a coarse-grained
pixelization of our default pixel scale ($nside=8192$ corresponding to
0.43~arcmin). Since our full-sky simulation is given in the hypothetical
RA and dec coordinates (see Section~\ref{subsec:mock}), we can assign
each random point of the redMaPPer catalog to the nearest {\tt HEALPix}
pixel. By identifying the pixels that contain at least one random point,
we define the survey footprints in the full-sky simulation. However,
note that our method does not account for the smaller-scale masks such
as those due to bright stars and does not consider other observational
effects such as variations in the survey depth that might affect the
selection of redMaPPer clusters in the real data.  The middle panel in
Fig.~\ref{fig:sdss_footprint} shows the effective area of the redMaPPer
catalog.  We then select the hypothetical redMaPPer clusters that reside
within the defined survey footprints in each simulation
realization. After these selection procedures, the total number of mock
clusters is typically greater than that of the real clusters (8,648) by
10--20\%, probably because we do not account for a redshift evolution in
the optical richness and halo mass relation and/or do not consider
observational effects such as the small-scale masks. Hence, we further
make a random downsampling of the mock clusters by the ratio of the
total numbers of mock and real clusters. Since the shear-cluster
cross-correlation probes the {\it average} mass distribution around the
sampled clusters, the estimator does not largely depend on the selection
of clusters or the total number of clusters in the sample
(equation~\ref{eq:est_delta_Sigma}).  Note that our approach of making a
mock catalog of redMaPPer clusters is different from the forward
modeling method, where the observed richness function is modeled by
convolving the theoretical halo mass function with the probability
function $P(\lambda|\ln M)$ that gives a probability distribution of
richness for a given halo mass $M$ including the scatters in $\lambda$
\citep{LimaHu:05,OguriTakada:11} (also see Murata et al. in
preparation). Since the main purpose of this paper is to study the
covariance for the stacked lensing, we believe that our mock catalog is
accurate enough, as we will show below in detail.

As we described, the source galaxies in the mock catalog (see
Section~\ref{subsec:sdss_shape}) include simulated lensing distortion
due to the mass distribution in the lens plane containing the mock
clusters. Hence, by taking the cross-correlation between the positions
of the mock redMaPPer clusters with the shapes of source galaxies, we
can perform a mock measurement of the stacked lensing. Note that we do
not consider an off-centering effect for simplicity; we use the true
halo center for the stacked lensing measurement in the mock catalogs, as
our main focus is to study the covariance at large scales. When
implementing the JK method in the mock catalog, we use the same
subdivision of survey footprints as in \citet{2016PhRvL.116d1301M},
including 63 and 20 for the northern and southern hemisphere footprints,
respectively (see the middle panel of Fig.~\ref{fig:sdss_footprint}).
The different realizations of the full maps contain different
realizations of mock clusters and large-scale structure, and allow us to
estimate the full covariance.

\subsubsection{Massive galaxies -- Luminous Red Galaxies}
\label{subsec:lrg}

We also consider another useful tracers of large-scale structure, the
sample of luminous red galaxies (LRGs)
\citep{2001AJ....122.2267E,Wakeetal:06}, which was constructed from the
SDSS DR7 data and made publicly available in
\citet{2010ApJ...710.1444K}.  The selection of LRGs are based on their
color and magnitude and well-designed to make the resulting sample be
approximately volume-limited out to $z\simeq 0.36$.  In this paper, we
use the same sample of LRGs as in \citet{2013MNRAS.432.1544M}, which
consists of 62,081 LRGs with $g$-band absolute magnitude $-23.2 < M_{g}
< -21.2$ in the redshift range of $0.16\le z < 0.36$. The LRGs are
early-type, bright galaxies and thought to reside in halos with a few
$10^{13}h^{-1}\, M_\odot$ that are less massive than the redMaPPer clusters.

In order to create the mock catalog of LRGs, we employ the halo
occupation distribution (HOD) method that allows us to populate
hypothetical LRGs into halos in the simulations.  The HOD, denoted by
$\langle N_{\rm gal}|M\rangle$, gives the mean number of galaxies in
host halos with mass $M$. Following the method in \citet{Zhengetal:05},
we employ the HOD model with the form given by
\beqa
\langle N_{\rm gal}|M\rangle 
&=& 
\langle N_{\rm cen}|M\rangle (1+\langle N_{\rm sat}|M\rangle), \label{eq:HOD_tot}
\eeqa
with
\beqa
\langle N_{\rm cen}|M\rangle 
&=&
\frac{1}{2}\left[ 1+{\rm erf}
\left(\frac{\log M-\log M_{\rm min}}{\sigma_{\log M}}\right)
\right], \label{eq:HOD_cen} \\
\langle N_{\rm sat}|M\rangle
&=& \left(\frac{M-M_{\rm cut}}{M_{1}}\right)^{\alpha}\Theta(M-M_{\rm cut}), \label{eq:HOD_sat}
\eeqa
where $\langle N_{\rm cen}|M\rangle$ denotes the HOD of central
galaxies, $\langle N_{\rm sat}|M\rangle$ is the contribution of
satellite galaxies and $\Theta(x)$ is the Heaviside step function.  In
this paper, we adopt the model parameters derived in
\citet{2009ApJ...698..143R}: $(M_{\rm min}, \sigma_{\log M}, M_{\rm
cut}, M_{1}, \alpha) =(5.64\times 10^{13}h^{-1}\, M_\odot,0.7, 3.50\times
10^{13}h^{-1}\, M_\odot, 3.47\times 10^{14}h^{-1}\,  M_\odot,1.035)$.  We do not
consider the redshift evolution of the HOD parameters.

Similarly to the redMaPPer catalog, we populate hypothetical LRGs in
halos in the range of $0.16\le z < 0.36$ for each realization of the
light cone simulations.
(1) We populate central LRGs into halos by randomly selecting halos
       according to the probability distribution, $\langle N_{\rm cen}|M
       \rangle$ (equation~\ref{eq:HOD_cen}).
       We assume that each central LRG resides at the halo center.
(2) For the halos that host the central LRG, we randomly populate
       satellite LRG(s) 
       assuming a Poisson distribution with the mean $\langle N_{\rm
       cen}|M\rangle$$\langle N_{\rm sat}|M\rangle$. We assume that the
       radial distribution of satellite LRGs on average follows that of dark
       matter in each host halo. Here we assume the analytical NFW profile
       \citep{Navarro:1996gj} for simplicity, where we use the halo mass
       and the scale radius, measured by the {\tt Rockstar}, to compute
       the NFW profile for each host halo. 
Then we compute the normalized radial distribution of mock LRGs in the
discrete 20 bins in the range of $0.16\le z < 0.36$. If the probability
of mock LRGs in a given bin is greater than that of the real LRGs, we
made a random downsampling of mock LRGs so as to match the radial
distribution. To make a mock measurement of galaxy-galaxy lensing for
the mock catalog of LRGs, we used 10 realizations of the random catalog.

Next, by comparing the random catalog with the simulated sky map in the
{\tt HEALPix} format of angular resolution parameter $nside=256$ 
(corresponding pixel size of about $13.6$ arcmin), 
we define the survey footprints of the LRG catalog.
The survey footprints of the LRG catalog is shown in the right panel of
Fig.~\ref{fig:sdss_footprint}. We then select the hypothetical LRGs in
each simulation that resides within the defined survey footprints.
After these procedures the total number of mock LRGs is found to be
close to the real number (62,081) with the scatter of $\sim100$. When
estimating the covariance in the JK method, we use the same subdivision of
the area into 100 subregions in \citet{2013MNRAS.432.1544M}, as
explicitly shown in the right panel of Fig.~\ref{fig:sdss_footprint}.

\section{RESULTS}\label{sec:result}
\subsection{Numerical tests of the covariance of stacked lensing}

Before going to the results for the mock catalogs of SDSS redMaPPer
clusters and LRGs, we first study the different covariance estimators of
stacked lensing profile using our light-cone simulations (lensing fields
as well as halo catalog), but not using the SDSS source galaxy catalog
for now.  For this purpose, we consider a sample of mass-limited halos
with $M\ge 10^{13.5}\, h^{-1}M_{\odot}$ and in the redshift range
$z=[0.10,0.27]$.  This sample typically contains $\sim 180,000$ halos
over one full-sky simulation. For the following results we use 10
realizations of the random catalogs. For source redshift, we simply
consider a single source plane at $z_{\rm s}=0.5$, distributed on the
regular grids (about 0.43~arcmin for grid scale), and will not consider
the intrinsic ellipticity or shape noise contamination to the covariance
unless otherwise stated.  The main purpose of the numerical experiments
is to understand the nature
of the JK covariance estimator in terms of the SSC
effect. For a reader who is more interested in the results for the SDSS
mock catalogs, please skip to Section~\ref{sec:sdss}.

\subsubsection{Comparison of the covariance estimators: full covariance
vs. jackknife covariance}
\label{subsec:basic_prop_cov}

We first compare the different covariance estimators using the
simulation realizations each of which has the same total
area, $\Omega_{\rm s}\simeq 859~$sq. degrees, corresponding to $f_{\rm
sky}=1/48$ for the sky fraction. Since our simulation maps are all given
in the {\tt HEALPix} pixelization, we define the hypothetical survey
footprint with $f_{\rm sky}=1/48$ by one pixel of the {\tt HEALPix}
pixelization with $nsize=2$\footnote{
See \url{http://healpix.sourceforge.net} for detail 
of the {\tt HEALPix} pixelization.
}. 
For comparison purpose, we use
exactly the same realizations of the hypothetical surveys when comparing
the different covariance estimators: we construct
480 realizations (each of which has $f_{\rm sky}=1/48$) from our
full-sky simulation realizations. To perform the JK covariance
estimation, we further subdivide each survey footprint by the finer
pixelization of {\tt HEALPix}, 12, 64, 256 subdivisions,
respectively. Table~\ref{tab:mass_limited_param} summarizes the survey
realizations, named as ``S859-X'', where ``X'' is either ``full'',
``jk'' or ``sub'' denoting either the full, jackknife or subsample
covariance estimator in Sections~\ref{sec:fullcov}, \ref{sec:jackknife},
or \ref{sec:subcov}, respectively.

\begin{table*}
\begin{tabular}{@{}llcccc|}
\hline
\hline
& Sky fraction $f_{\rm sky}$ & Number of subregions & Number of realizations \\ \hline
 S859-full
 & $f_{\rm sky}=1/48$ (859.4 sq.~degs) & -- & 480 \\
 S859-jk1/sub1 & $f_{\rm sky}=1/48$ (859.4 sq.~degs) & 12 & 480 \\
 S859-jk2/sub2
 & $f_{\rm sky}=1/48$ (859.4 sq.~degs) 
	 & 64 & 480 \\
S859-jk3/sub3 & $f_{\rm sky}=1/48$ (859.4 sq.~degs) 
& 256 & 480 \\
S3437-full & $f_{\rm sky}=1/12$ (3437 sq.~degs) 
& -- & 120 \\
S215-full & $f_{\rm sky}=1/192$ (214.8 sq.~degs) & -- & 1920 \\
S54-full & $f_{\rm sky}=1/768$ (53.71 sq.~degs) & -- & 7680 \\
\hline
\end{tabular}
 \caption{Summary of the parameters of simulations used for the
 comparison of different covariance estimators for numerical experiments
 of galaxy-galaxy weak lensing for a sample of mass-selected halos with
 $M_{\rm 200m}\ge 10^{13.5}\, h^{-1}\, M_\odot$ and in the redshift range $z_{\rm
 l}=[0.1,0.27]$. The different columns denote the sky coverage ($f_{\rm
 sky}$) of each simulation realization, whether the covariance is
 estimated with or without subdivision of each realization, and the
 number of realizations, for each experiment. 
 } \label{tab:mass_limited_param}
\end{table*}

\begin{figure*}
\centering \includegraphics[width=0.5\columnwidth, bb=0 0 523 593]
 {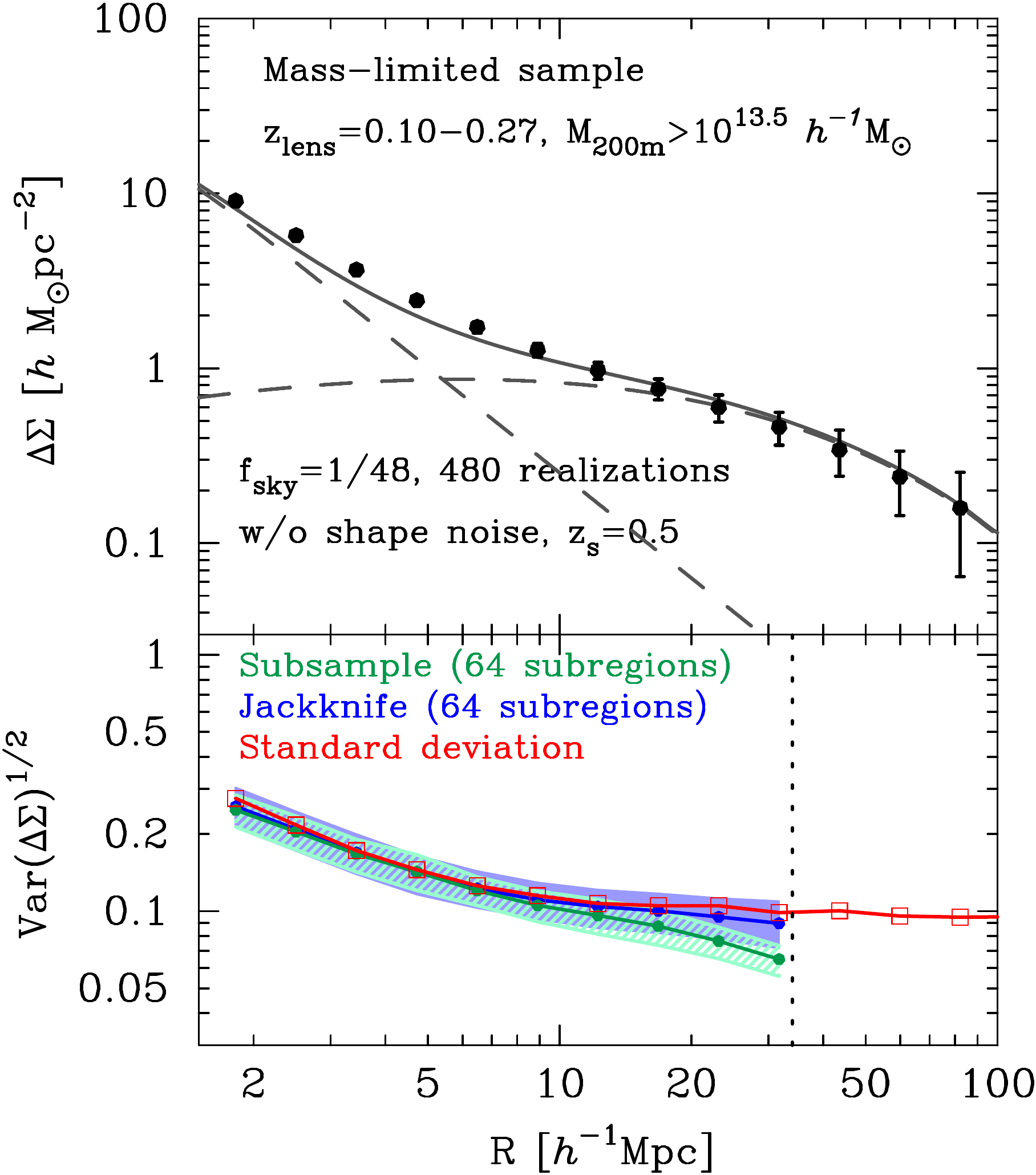}
 \caption{A simulation of the galaxy-galaxy weak lensing for a
hypothetical survey with survey area $\Omega_{\rm s}\simeq
859~$sq. degrees corresponding to the sky fraction of $f_{\rm
sky}=1/48$. We consider, as lensing objects, a sample of mass-limited
halos with $M_{\rm 200m}\ge 10^{13.5}\, h^{-1}M_{\odot}$ in the redshift range of $0.10-0.27$ (see text for details).  
As for source galaxies we assume that all the 
galaxies are at a single redshift $z_{\rm s}=0.5$ for simplicity, 
and are regularly distribution in grids spaced by 0.43~arcmin
 (corresponding to the number density $\bar{n}_{\rm s}\simeq
 5.4~$arcmin$^{-2}$).  We do not consider the intrinsic shapes. 
 Note that 
 the simulated lensing effect on each source galaxy includes not only
 the mass contribution in the lens planes, but also the mass
 contribution along the line of sight at different redshifts, i.e. the
 projection effect. For this study we use the 480 realizations of mock
 simulations (halo catalogs and source galaxies).  {\em Upper panel}:
 The black points show the average signal over the 480 realizations,
 while the error bars are estimated from diagonal components of the full
 covariance matrix, i.e. the standard deviation of the 480
 measurements. For comparison, the solid and dashed curves show the
 analytical, halo model predictions (see text for details).  {\em Lower
 panel}: Comparison of the different covariance estimators, the
 jackknife, subsample and full covariance methods, in
 Section~\ref{sec:cov}.  The open square symbols, connected by the red
 line, show the diagonal components of the full covariance at
 each separation bin. On the other hand, the blue and green circles are
 the results for the jackknife and subsample methods; the covariance of
 each method is estimated from each realization, the circles are the
 average of the 480 covariance matrices, and the shaded regions around
 the curve show the 10-90\% range of distribution in the
 480 measurements. 
 All the covariance estimations 
 include the subtraction of the random signals. 
 The vertical dotted line denotes the projected length corresponding to
 the size of each subregion at the lens redshift. For these results we
 used 10 realizations of the random catalogs of lensing halos for each
 of the lens map realizations (see equation~\ref{eq:est_delta_Sigma}).
 } \label{fig:dSigma_masslimited_fsky48}
\end{figure*} 

The upper panel of Fig.~\ref{fig:dSigma_masslimited_fsky48} shows the
surface excess mass density profile, $\Delta\Sigma(R)$, computed from
the average of 480 realizations of the mock catalog of mass-limited
halos with $M_{\rm 200m}\ge 10^{13.5}~{h}^{-1}M_\odot$.  We employ 15
equally-spaced logarithmic bins with the bin width of $\Delta \ln
R=0.31$ in the range of $1.5<R \, [h^{-1}\, {\rm Mpc}]<178$. 
The expectation value of stacked lensing probes the average mass
distribution around the lensing halos as a function of centric radii,
and we confirmed that the sufficient number of realizations gives a
well-converged value at each radial bin. 
For comparison, the solid curve is the analytical halo model prediction, where we employed the same
model ingredients in \citet{2013MNRAS.435.2345H}; the model ingredients
are the NFW profile for the halo mass density profile and the fitting
formulas for the mass function and the halo bias
\citep{2008ApJ...688..709T,2010ApJ...724..878T}.  
The analytical model shows a 10-20\%-level agreement with 
the simulation results over the range of radii. The agreement validates
an accuracy of the $N$-body simulations and halo catalogs we are using
in this paper.  The error bars at each bin are estimated from the
diagonal components of the full covariance, estimated from the standard
deviation of 480 realizations, denoted as ``S859-full'' in
Table~\ref{tab:mass_limited_param}. Although we did not include the
shape noise, the error covariance arises from various contributions: a
shot noise due to a finite number of lensing halos, the sample variance
of mass distribution in the lens planes, the projection effect of
large-scale structure along the same line of sight out to sources at
$z_{\rm s}=0.5$, and the SSC contribution.

The lower panel compares the different covariances estimated based on
the full covariance method (square), the JK method (blue circle) or the
subsample method (green circle). Note that we used exactly the same 480
realizations of mock catalogs to estimate the different covariances. For
the JK and subsample methods, we estimate the covariance matrix in each
realization. The bold lines connected by the circle symbols are the
average of 480 covariance matrices, and the shaded region around the
lines show the 10-90\% distribution in the 480 measurements. We
confirmed that the subtraction of the signals around random points (the
2nd term in equation~\ref{eq:est_delta_Sigma}) is important for the
covariance estimation, as claimed in \citet{2016arXiv161100752S}. While
including the 2nd term in the estimator changes the signal by up to a
few 10\% in the amplitude at the large scales up to $\sim 100~$Mpc$/h$,
it reduces the variance amplitudes by up to a factor of 2. The amount of
the reduce by the random subtraction is larger in the order of the
subsample, JK and full covariance methods. The random subtraction
corrects for the cosmic shear contribution due to a general survey
geometry, as implied by the last term of equation~(10) in
\citet{2016arXiv161100752S}; the random correction is not needed for the
halo-matter cross-correlation estimation if a geometry obeys periodic
boundary conditions, e.g. as in $N$-body simulation box, although it is
not realized in an actual observation. We should note that, if we do not
employ the random subtraction, the variances for the JK and subsample
methods are larger than that of the full covariance method.

The lower panel shows that, at radii $R\simlt 6~h^{-1}\, {\rm Mpc}$, all
the covariance matrices agree with each other on average, implying that
each method gives an unbiased estimation of the error bars.  
The error bars at the small separations arise
mainly from the Gaussian term (and the non-Gaussian term of sub-survey
modes, although seems subdominant),
because the term scales with the
number of pairs used in the stacked lensing measurement at each
separation bin; $[{\rm Var}(\Delta\Sigma)]^{1/2}\propto 1/\sqrt{N_{\rm
pair}}\propto 1/[\Omega_{\rm s}R^2]^{1/2}$ for logarithmically spacing
bins.  At the larger separations $R\simgt 6h^{-1}\, {\rm Mpc}$, the JK
method is in good agreement with the full
covariance method at a level of 10\% in the amplitude,
 while the subsample method appears to under-estimate the error. 
We should note that the JK and subsample variances have a scatter and
are therefore noisy in each realization basis.

\begin{figure*}
\centering
\includegraphics[width=0.6\columnwidth, bb = 0 0 562 404]
 {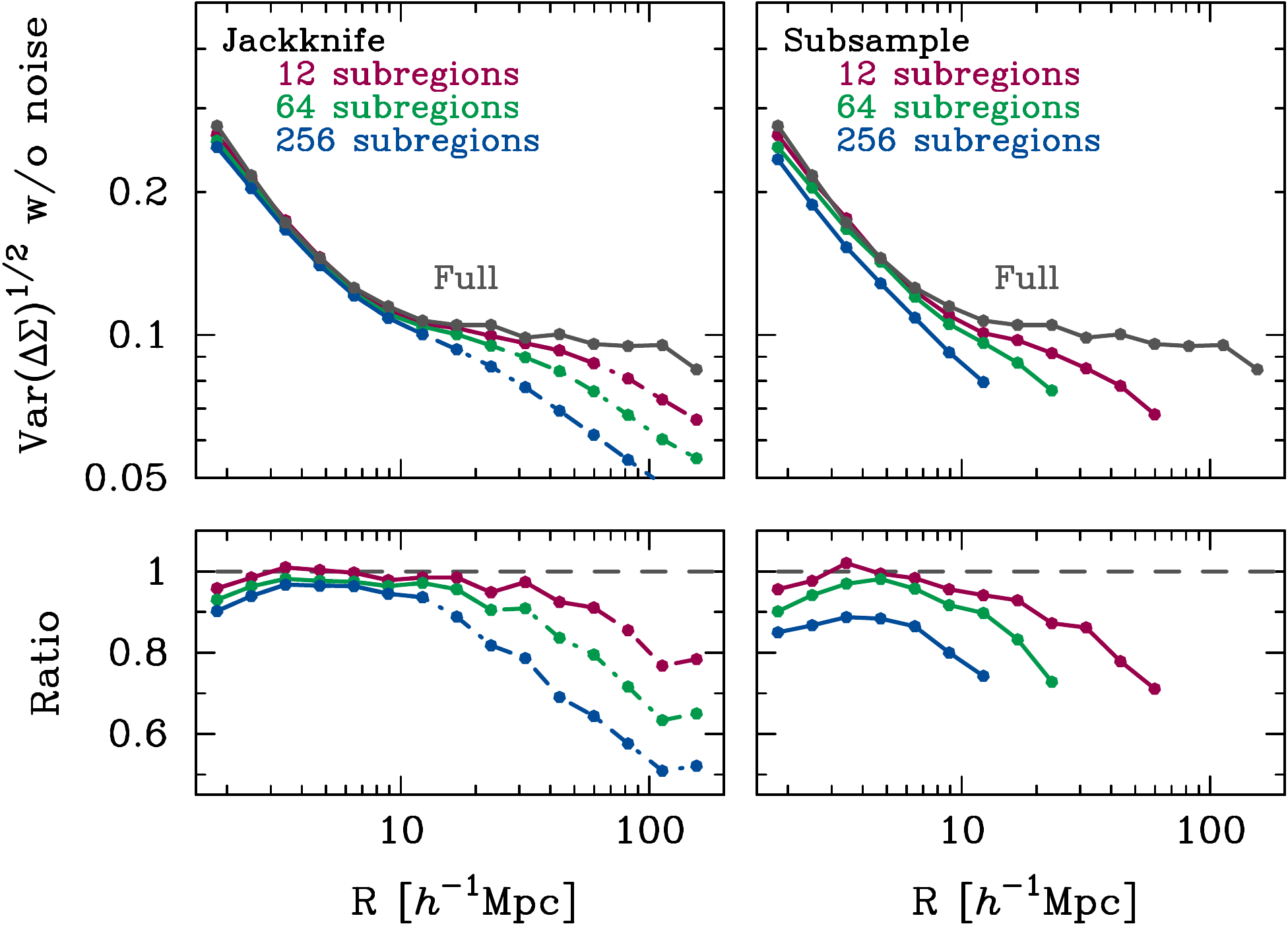}
 \caption{ {\em Left panels}: The dependence of the jackknife
 covariances on different subdivisions. For comparison we used the same
 480 realizations as in Fig.~\ref{fig:dSigma_masslimited_fsky48}, each
 of which has an area of $859~$sq. degrees ($f_{\rm
 sky}=1/48$). We make different subdivisions, 12, 64, or 256 subregions,
 for each realization (as denoted by S859-X in
 Table~\ref{tab:mass_limited_param}), and then perform the jackknife
 covariance estimation for each case. The circle symbols show the
 average of the 480 covariance matrices. For comparison, we also show
 the full covariance matrix.  The lower panel shows the ratio of each
 jackknife covariance relative to the full covariance. 
 The jackknife curve traces the full covariance within the corresponding separation of the projected size of jackknife subregion for the mean lens redshift ($\bar{z}_{\rm l}\simeq 0.18$), and then starts to decrease at the larger separations as shown by the dashed curve.
 {\em Right}: Similarly to the left panels, but for the
 subsample covariance method. Here we plot the result up to the scale
 corresponding to the projected size of subregion for the mean lens
 redshift. 
 }
 \label{fig:var_dSigma_Nsub}
\end{figure*} 

We also compare the JK methods of
different subdivisions for a fixed total area ($f_{\rm sky}=1/48$) in
the left panel of Fig.~\ref{fig:var_dSigma_Nsub}. Here we consider the
JK results for 12, 64 and 256 subdivisions corresponding to 72, 13, and
3.4~sq. degrees for each subregion, respectively.
Again notice that we used exactly the same 480 realizations for the
covariance estimations. 
To compute the JK covariance we remove 
only the lensing halos in each subregion for the JK
resampling as done in the actual measurements
\citep[e.g.,][]{2013MNRAS.432.1544M}. 

Fig.~\ref{fig:var_dSigma_Nsub} shows that all the covariance estimators
including the full covariance are in fairly nice agreement with each
other at small separations on average, as we found in
Fig.~\ref{fig:dSigma_masslimited_fsky48}.  
The JK variances are in nice agreement with the full variance at radii
up to the size of JK subregion, to within about 10\% in the amplitude
for different subdivisions.
At the larger separations, the curve starts to decrease, 
i.e. under-estimating the true covariance, by a larger amount at the larger radii. 
This is because the stacked lensing measurements at
separations greater than the JK subregion are in less variations between
the different JK resamples for a given realization.
In Appendix~\ref{app:off_diag_dSigma_Nsub}, we show the off-diagonal
terms of JK covariance as in Figure~\ref{fig:var_dSigma_Nsub}.  We found
that the off-diagonal term is less sensitive to the number of JK
subregions once it is normalized by the diagonal term.

The right panel of Fig.~\ref{fig:var_dSigma_Nsub} shows the covariance
matrices for the subsample method using the same subdivisions of total
area as in the JK method in the left panel. We again notice that, in
this method, we use different realizations of the light-cone simulations
to generate the lensing fields and lensing halos in each subregion, and
there is no correlation between the lensing fields in different
subregions.  
Except for the amplitude,  
the subsample method shows a similar scale dependence of the covariance 
to what is found in the JK method up to the separation scale corresponding 
to the projected size of subregion.  
The overall difference among three cases in the right panel 
can be partly explained by the averaging effect of the subsample variance
over different realizations.
Since the variance at small radii should be scaled with $1/N_{\rm pair}$,
the averaged subsample variance over realizations induces the difference
between $\langle1/N_{\rm pair}\rangle$ and $1/\langle N_{\rm pair}\rangle$.
The factor of $\Omega_{\rm sub}/\Omega_{\rm s}$ in equation~(\ref{eq:cov_sub})
is intended to correct for the difference of $\langle N_{\rm pair} \rangle$ between the area 
of total survey region and the subregion.
We find the difference 
between $\langle1/N_{\rm pair}\rangle$ and $1/\langle N_{\rm pair}\rangle$ 
over 480 realizations
by 3\%, 10\% and 30\% for 12, 48, and 256 subdivisions respectively.
Once we include these differences, the subsample method can show a
reasonably good fit to the true covariance at $R<10\, h^{-1} {\rm Mpc}$.
The tendency of $\langle1/N_{\rm pair}\rangle \langle N_{\rm
pair}\rangle \neq 1$ should be more prominent for a smaller sky
coverage. We confirmed that the difference arises from the sample
variance in the number of lensing halos, a sort of the super sample
variance effect, as first pointed out in \citet{2003ApJ...584..702H}.

\subsubsection{Effect of shape noises}
\label{subsubsec:shape_noise}
\begin{figure*}
\centering
\includegraphics[width=0.6\columnwidth, bb=0 0 562 410]
 {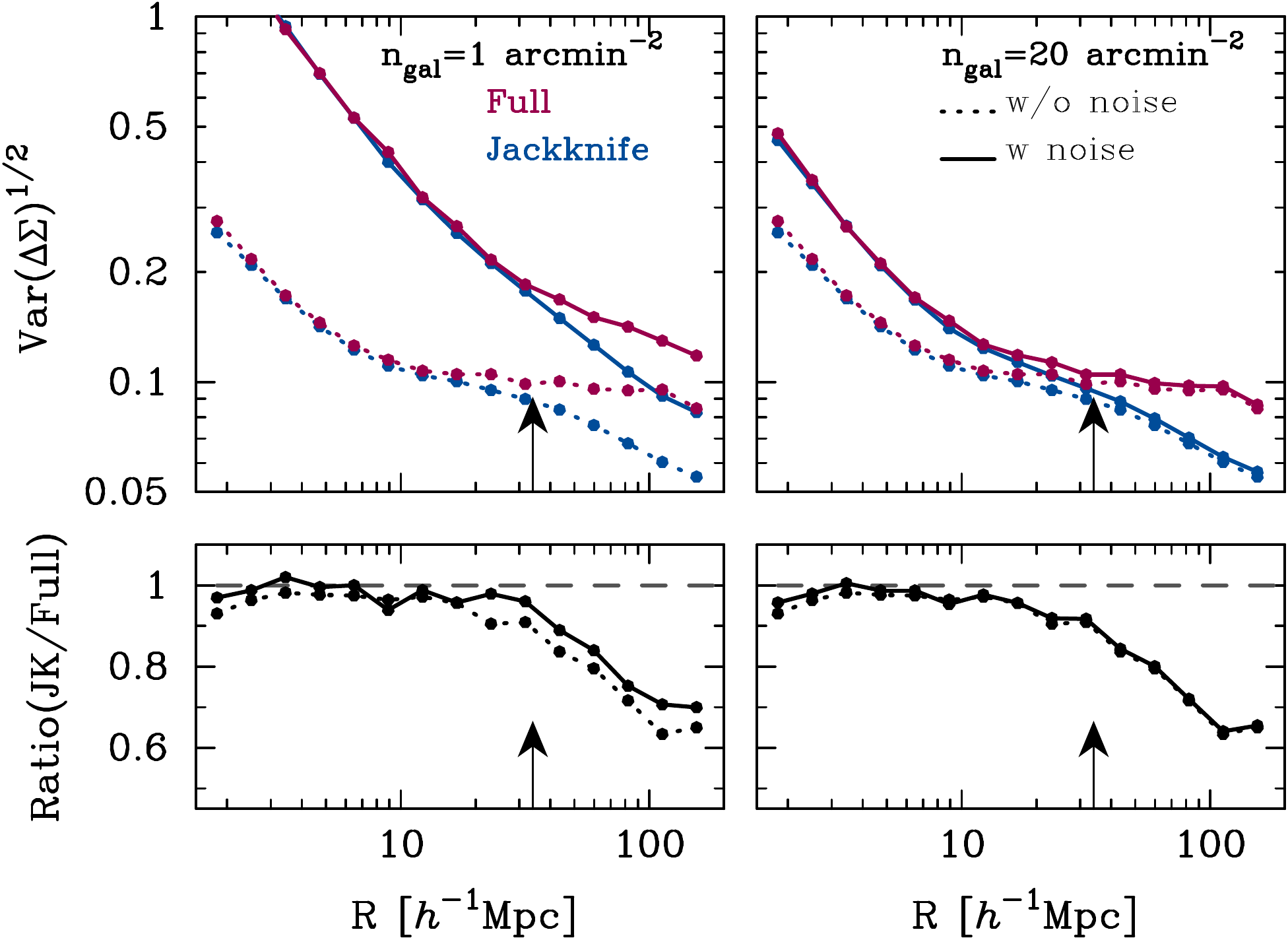}
\caption{
The effect of shape noises in the covariance estimations.
In this figure, we used the same 480 realizations as in 
Fig.~\ref{fig:dSigma_masslimited_fsky48} with 64 subdivisions for each realization. When performing the covariance estimation, 
we include the shape noise contribution by adding the Gaussian noise 
with a fixed rms of shape ellipticities of 0.35, but different number density of sources is assumed. The arrow in each panel corresponds to the projected size of JK subregion for the mean lens redshift ($\bar{z}_{\rm l}\simeq 0.18$).
{\em Left panels}: The case of source number density of 
1~${\rm arcmin}^{-2}$, 
corresponding to the current survey such as the SDSS. 
The blue and red lines show the JK and the full variance, respectively.
For each covariance, the solid line represents the presence of shape noises,
while the dashed line is for the absence of shape noises.
The lower panel shows the ratio of each JK covariance 
relative to the full covariance. 
 {\em Right}: Similarly to the left panels, but for the
 case of source number density of 20~${\rm arcmin}^{-2}$.
 }
\label{fig:var_dSigma_fsky48_noise}
\end{figure*} 

We then examine the impact of shape noises on the comparison between the
JK and full covariance methods.  We use the 480 realizations as denoted
by S859-jk2 in Table~\ref{tab:mass_limited_param}.
We add, to the simulated shear at each pixel in the source plane
($z_s=1$), a random ellipticity assuming the Gaussian distribution given
by
\beqa
P(e) = \frac{1}{\pi\sigma_{\rm int}^{2}}
\exp\left(-\frac{e^2}{\sigma_{\rm int}^2}\right),
\eeqa
where $e^2=e_1^2+e_2^2$ and $\sigma_{\rm int}^2 = \sigma_{e}^2/(n_{\rm
gal}\theta_{\rm pix}^2)$ with the pixel size of 0.43 arcmin.  We then
assume a random orientation to determine the two elliptiicity
components. Here, we set $\sigma_{e}=0.35$ and study two cases of
$n_{\rm gal} = 1$ or $20\, {\rm arcmin}^{-2}$.  The former source number
density corresponds to a typical value of the imaging survey in SDSS,
while the latter is for imaging surveys such as the Subaru Hyper
Suprime-Cam survey.

Fig.~\ref{fig:var_dSigma_fsky48_noise} compares the JK and full
variances in the presence of shape noises.  In the case of $n_{\rm
gal}=1\,{\rm arcmin}^{-2}$ where the shape noise gives a greater
contribution to the variance
over the wider range of $R$, the JK method gives a reasonably good
agreement with 
the full variance up to the projected size of JK subregion at the mean
lens redshift.  In the case of
$n_{\rm gal}=20\,{\rm arcmin}^{-2}$ where the sample variance gives a greater contribution relative to the shape noise,
the JK variance gives a good agreement with the full variance in the
shape noise dominated regime, $R\simlt 10~{\rm Mpc}/h$ for this case,
but shows about 10\% underestimation in the error (20\% in the variance)
at the radius corresponding to the JK subregion size (denoted by the
arrow in the $x$-axis). The underestimation becomes greater at radii
beyond the JK subregion size.

\subsubsection{Area dependence of SSC in the full covariance}
\label{subsec:area_dependence_full_cov}

\begin{figure*}
\centering
\includegraphics[width=0.5\columnwidth, bb=0 0 457 474]
 {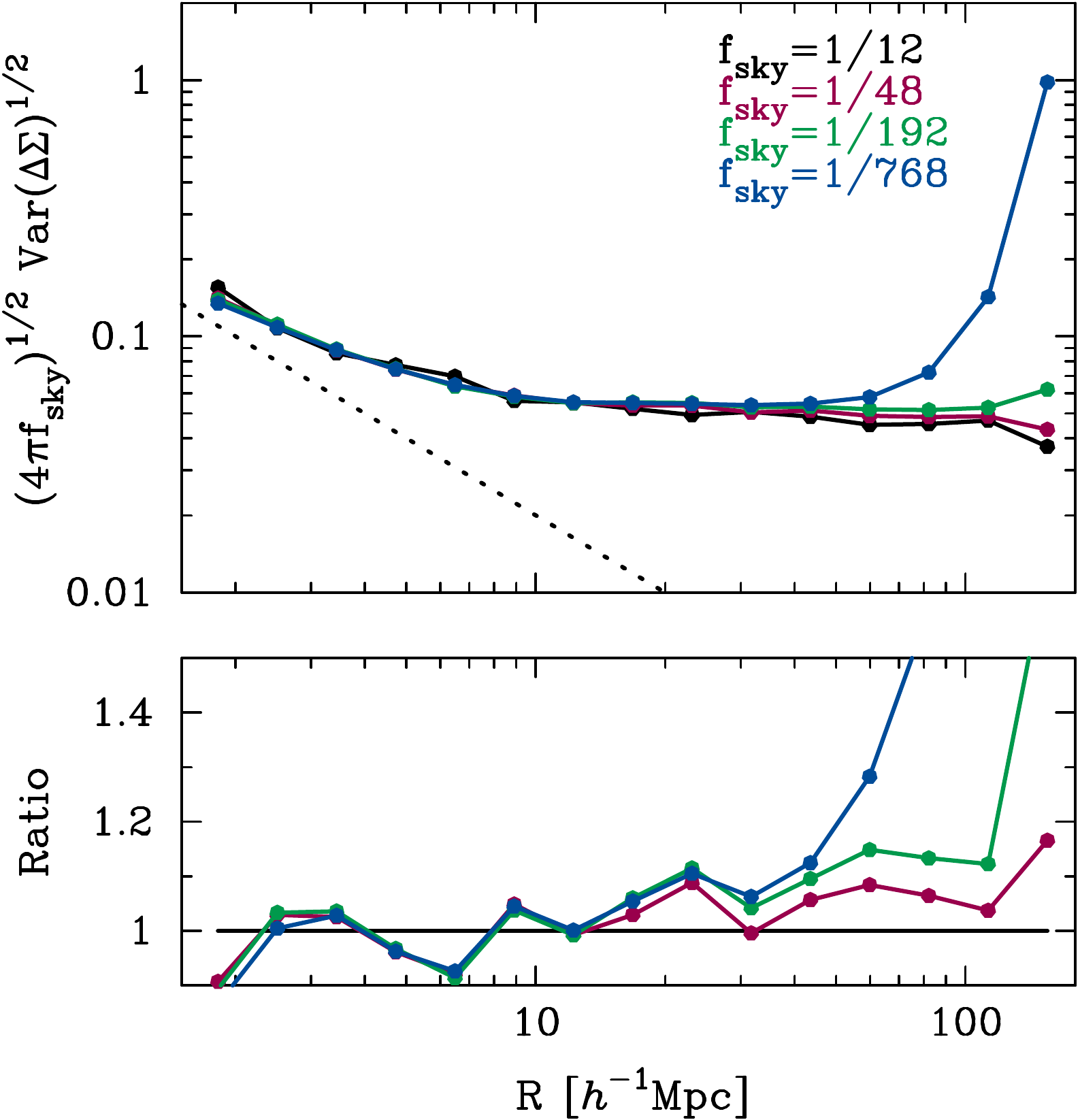}
\caption{ The dependence of the full covariance on the sky coverage,
$f_{\rm sky}=1/12, 1/48, 1/192$ or $1/768$, respectively. We used
exactly the same realizations of different numerical measurements. We
plot the covariance matrix for each case, multiplied by a factor of
$(4\pi f_{\rm sky})^{1/2}$, such that all the covariance matrices agree
with each other if the covariance matrix scales with the sky fraction as
${\bf C}\propto 1/f_{\rm sky}$. If the covariance scales with the
inverse of the number of source-lens pairs, it should scale with $R$ as
${\bf C}\propto 1/{N_{\rm pair}}\propto [R^2\Delta \ln R]^{-1}$ for the
logarithmically spaced bins, as denoted by the dotted line. In the
bottom panel, we show the ratio of the different covariance matrices
relative to the covariance for the largest area case ($f_{\rm
sky}=1/12$), which has a minimum impact of the SSC term (see text for
details).  } \label{fig:var_dSigma_fsky}
\end{figure*} 

Is the variance of galaxy-galaxy weak lensing 
is sensitive to the SSC term?  
Fig.~\ref{fig:var_dSigma_fsky}
addresses this question, showing how the diagonal components of the full
covariance scale with the survey area. Here we consider the light-cone
simulations for different areas of $f_{\rm sky}=1/12, 1/48, 1/192$ and
$1/768$ corresponding to $\Omega_{\rm s}\simeq 3437, 859, 215$ and $54$
sq. degrees, respectively, for the mass-limited sample of lensing halos
with $M_{\rm 200m}\ge10^{13.5}~h^{-1}\, M_\odot$ as used in
Figs.~\ref{fig:dSigma_masslimited_fsky48} and
\ref{fig:var_dSigma_Nsub}. These simulations are given by ``SXXX-full''
in Table~\ref{tab:mass_limited_param}. The survey footprints for these
simulations are defined by the {\tt HEALPix} pixelization. The footprint
of each realization for the largest area, $f_{\rm sky}=1/12$, is given
by one {\tt HEALPix} pixel, set by the parameter $nside=1$. Then the
smaller area realizations are defined by further, hierarchical {\tt
HEALPix} pixelizations of each region of $f_{\rm sky}=1/12$. By doing in
this way, we use exactly the same light-cone simulations for this
comparison study.  Hence the difference in the covariance should arise
from the SSC term, i.e. the effect of super-survey modes for each survey
footprint.

Fig.~\ref{fig:var_dSigma_fsky} shows the diagonal component of the
full covariance, multiplied by $(4\pi f_{\rm sky})^{1/2}$. If the
covariance scales with $1/f_{\rm sky}$, all the curves should coincide,
which is indeed the case at small separations, $R\simlt 10~h^{-1}\, {\rm
Mpc}$. In this regime, the error bars scale with the number of pairs
between source and lensing halos, leading to $[{\rm
Var}(\Delta\Sigma)]^{1/2}\propto [1/R^2\Delta\ln R]^{1/2}\propto 1/R$
for the fixed logarithmic binning of $R$, as denoted by the dotted line.
At the larger separations, the scaling of covariance with $1/f_{\rm
sky}$ breaks down.  The full covariance with smaller total area starts
to deviate from the covariance of $f_{\rm sky}=1/12$ at a smaller
separation, because it has a greater contribution of the SSC term.
Together with the results in Fig.~\ref{fig:var_dSigma_Nsub}, one might
think that, by measuring the JK covariances as a function of different
subdivisions, we can infer the full covariance for the total area of a
given survey by scaling the JK covariance according to the area
dependence of the SSC term shown in
Fig.~\ref{fig:var_dSigma_fsky}. However, this would be difficult in
practice because the JK covariance estimation can be noisy for a
particular one realization (see below), especially when the dimension of
data vector is large, and also because the SSC term depends on the
geometry of survey footprints, and therefore cannot be specified by the
single quantity $f_{\rm sky}$ even if the underlying cosmological model
is known \citep{2013PhRvD..87l3504T,Takahashietal:14}. 
From the results in
Figs.~\ref{fig:dSigma_masslimited_fsky48}--\ref{fig:var_dSigma_fsky},
we conclude that the SSC term is a subdominant source of the sample variance 
at $R\simlt30~h^{-1}{\rm Mpc}$, 
corresponding to the typical largest scale in the current stacked lensing analyses,
while the SSC term can be significant
at $R\sim100~h^{-1}\, {\rm Mpc}$.
In Appendix~\ref{app:off_diag_dSigma_fsky}, 
we show off-diagonal terms of the full covariance matrix as in Figure~\ref{fig:var_dSigma_fsky}.

\subsection{Application to SDSS data: redMaPPer clusters and LRGs}
\label{sec:sdss}

\begin{figure*}
\centering
\includegraphics[width=0.45\columnwidth, bb=0 0 493 573]
{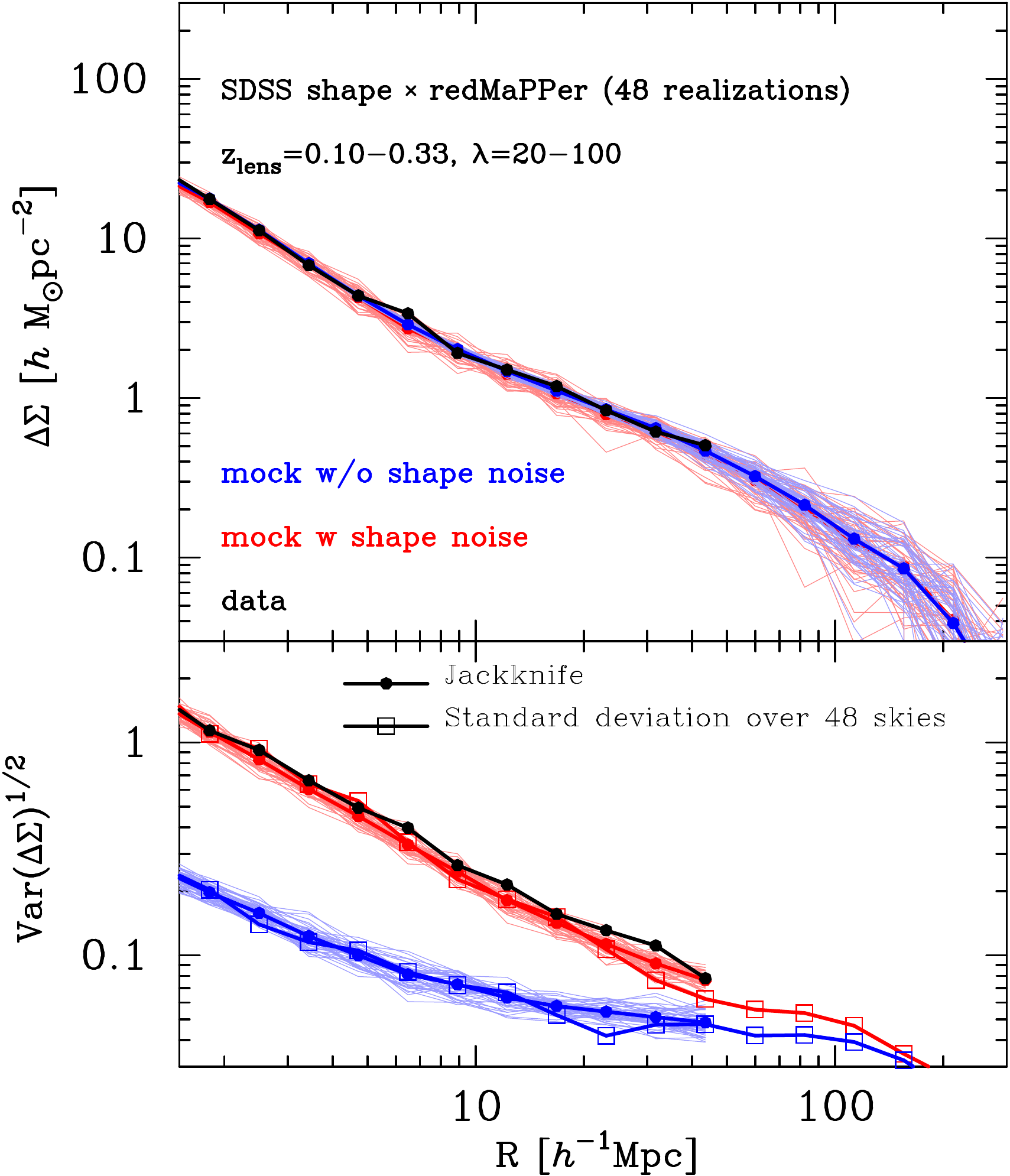}
\includegraphics[width=0.45\columnwidth, bb=0 0 493 573]
{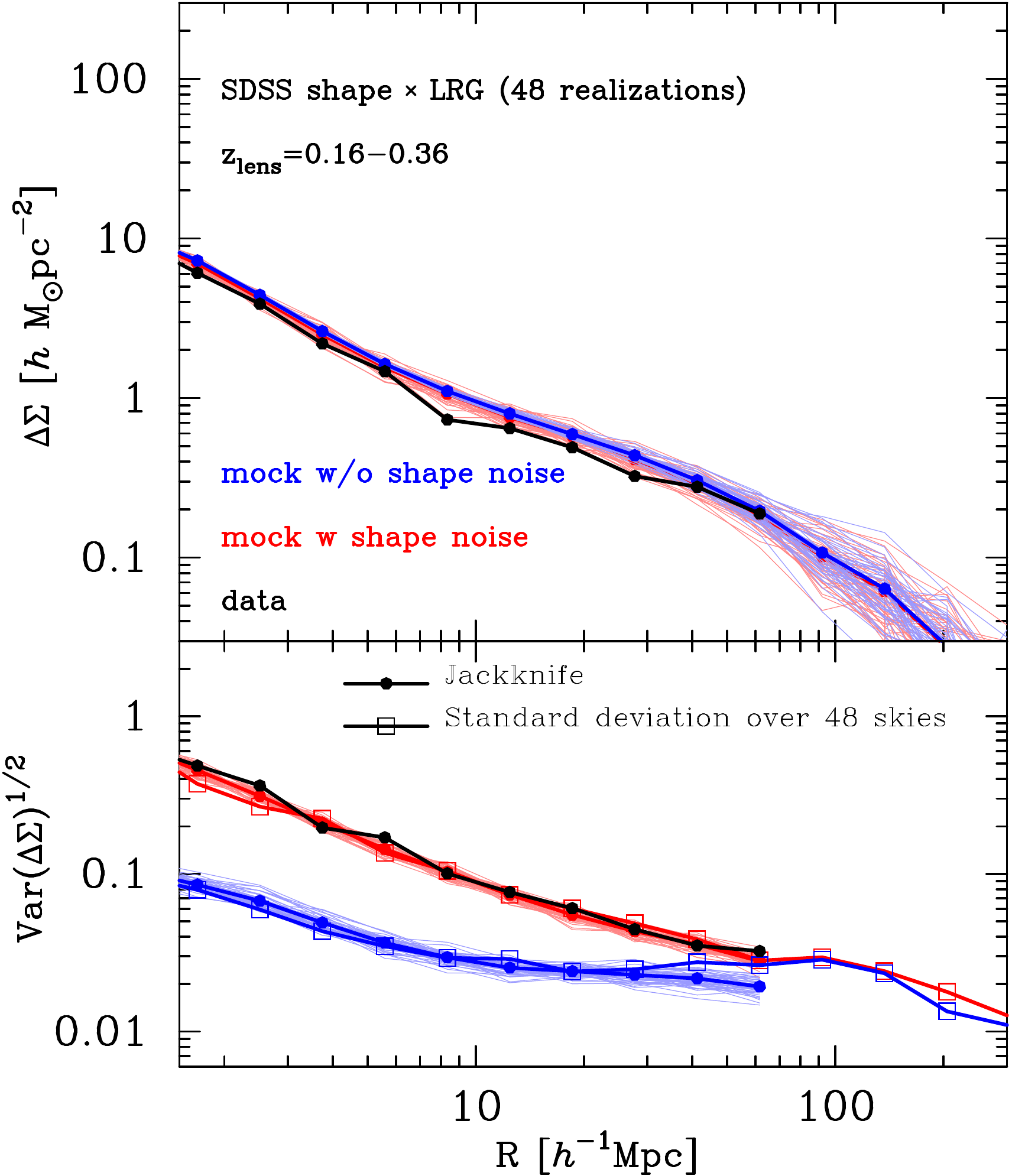}
 \caption{
Comparison of the stacked signals measured from the mock
catalogs and the real SDSS data, for the redMaPPer clusters (left
panels) and LRGs (right), respectively.  To do this, we used the 48
realizations of full-sky, light-cone simulations for the $\Lambda$CDM
cosmology; we populate the real SDSS source galaxies (their sky
distribution and intrinsic shapes), simulate the lensing effects on each
source galaxy and identify hypothetical redMaPPer clusters or LRGs in
the halo catalogs (see Section~\ref{sec:sim}).
We also properly take
into account the footprints for both the source galaxies and the lensing
objects in Fig.~\ref{fig:sdss_footprint}.  In each panel, the upper
plot shows the stacked lensing profiles, where we employ the same
logarithmically-spacing binning as used in
\citet{2016PhRvL.116d1301M}. The black circle symbols are the
 measurements from the SDSS data;
we used the results of 
 \citet{2016PhRvL.116d1301M} for the redMaPPer clusters and
re-performed the LRG weak lensing measurements for $\Delta\Sigma(R)$ following
\citet{2013MNRAS.432.1544M} (see text for details),
respectively. The red points are the average of the 48 realizations,
while the thin curve shows the result for each realization. For
comparison, the blue point and the thin curve show the results without
shape noise.  The lower plot shows the diagonal components of the
covariance matrix for the full covariance method or the jackknife
method, where we used the same jackknife subdivision in
Fig.~\ref{fig:sdss_footprint} as in the measurements. The black points
are the jackknife covariance estimated from the real SDSS data. Our
light-cone simulations fairly well reproduce the signals and JK
covariances from the real SDSS data.
  } \label{fig:sdss_dSigma}
\end{figure*} 

Now we move to an application of our method to the real SDSS data, the
catalogs of redMaPPer clusters and LRGs. We again emphasize that we
utilize 48 realizations of full-sky simulations as described in
Section~\ref{sec:sim}: we populate the real catalog of SDSS source
galaxies into the source planes, define the survey footprints, identify
hypothetical lensing objects in each lens plane and simulate the lensing
effect on each source galaxy.  For each mock catalog, we measured the
stacked lensing profiles. We then implemented the covariance
estimations: we combined all the 48 realizations to estimate the full
covariance, while we performed the JK covariance estimation for each
realization. Note that, for this study, we used the JK resampling method
to remove only the lensing objects in each subregion, but use all the
source galaxies for each JK resample, as used in the measurements.

The left panels of Fig.~\ref{fig:sdss_dSigma} show the result for our
mock realizations of the redMaPPer clusters. We used the same
logarithmically-spacing binning of radial separations as used in
\citet{2016PhRvL.116d1301M}. The upper plot shows that our mock catalogs
well reproduce the stacked lensing profile measured from the real SDSS
data over the range of radii we considered.  The lower plot compares the
diagonal components of covariance matrix at each radial bin, measured
from the mock catalog and also from the real data.  The thin red curves
show the covariance matrix at each radial bin, measured from each
realization based on the JK method.  The bold red circle shows the
average of 48 JK covariance matrices, and can be compared with the JK
covariance measured from the real SDSS data in
\citet{2016PhRvL.116d1301M}. It can be found that our mock catalog
fairly well reproduces the JK covariance of the real data.  The square
symbol shows the full covariance, which matches the JK errors at
$R\simlt 50-60~h^{-1}\, {\rm Mpc}$.  These results show that the JK
estimator gives an unbiased estimator of the covariances when the
separation scale is smaller than the size of JK region. We again note
that the agreement is realized, if the random catalog subtraction is
implemented as claimed in \citet{2016arXiv161100752S}.
When the random subtraction is not implemented, we find that the JK
method over-estimates the errors at $R\simgt 10-20~h^{-1}\, {\rm Mpc}$.
For comparison, the blue-thin curves and -circle symbols show the full
or JK covariances without the shape noise contamination. Comparing the
red and blue curves manifests that the covariances at $R\simgt
50~h^{-1}\, {\rm Mpc}$ are in the sample variance regime for the SDSS
number density of source galaxies ($\bar{n}_{\rm gal}\sim
1~$arcmin$^{-2}$).  For the full covariance case, we can use the stacked
lensing measurements up to the greater separations than in the JK
method, where we can use the measurements only up to the maximum
separation corresponding to the size of JK subregion.

The right panel of Fig.~\ref{fig:sdss_dSigma} shows the similar results
for the SDSS LRGs. Note that \citet{2013MNRAS.432.1544M} used the
compensated weak lensing profile $\Upsilon(R;R_0)$, which is designed to
remove the small-scale information at $R\simlt 4~{\rm Mpc}/h$, for the
cosmological analysis. They carefully discussed the covariance matrix
for the $\Upsilon$-profile, and found that it appears to be less
sensitive to the non-Gaussian sample variance due to the nature of the
statistic. In this paper, we re-performed the measurements of
$\Delta\Sigma(R)$ for the same LRG sample, and estimated the JK
covariance following the JK subdivision in \citet{2013MNRAS.432.1544M}.
The agreements between the JK covariances from our mock catalogs and the
real data are somewhat better, probably due to the higher number density
of LRGs than that of the redMaPPer clusters. 

In Appendix~\ref{app:off-diagonal} we study the off-diagonal components
of the covariance matrix, and found that the JK covariance, estimated
from one particular realization, can be noisy and cannot reliably
estimate the off-diagonal components, if an insufficient number of JK
subregions are used. 

\subsection{Implications for cosmological analysis}

\subsubsection{An improvement in the signal-to-noise ratio due to the
use of accurate covariance}

As we have so far shown, the JK covariance is noisy on each realization
basis, although it gives a reasonably accurate estimation of the true
covariance on average. In addition, the JK covariance is valid only up
to the size of JK subregion.
Therefore the use of an accurate covariance in cosmological analysis
should become important to obtain robust results. 
(1) The JK covariance can be noisy, because the covariance is estimated
from a particular realization, i.e. the real data, and if the number of
the JK subregions is not sufficient.  (2) Since the JK method sets a
maximum separation scale corresponding to the size of JK subregion in
the cosmological analysis, the use of the full covariance, estimated
from the mock catalogs, gives us an access to the larger separations,
where there is a cleaner cosmological information in the weakly
nonlinear or linear regime.  (3) The JK method tends to under-estimate
the true covariance at the boundary of JK subregion for a survey with
higher number density of source galaxies. 

\begin{figure*}
\centering
\includegraphics[width=0.45\columnwidth, bb=0 0 435 465]
{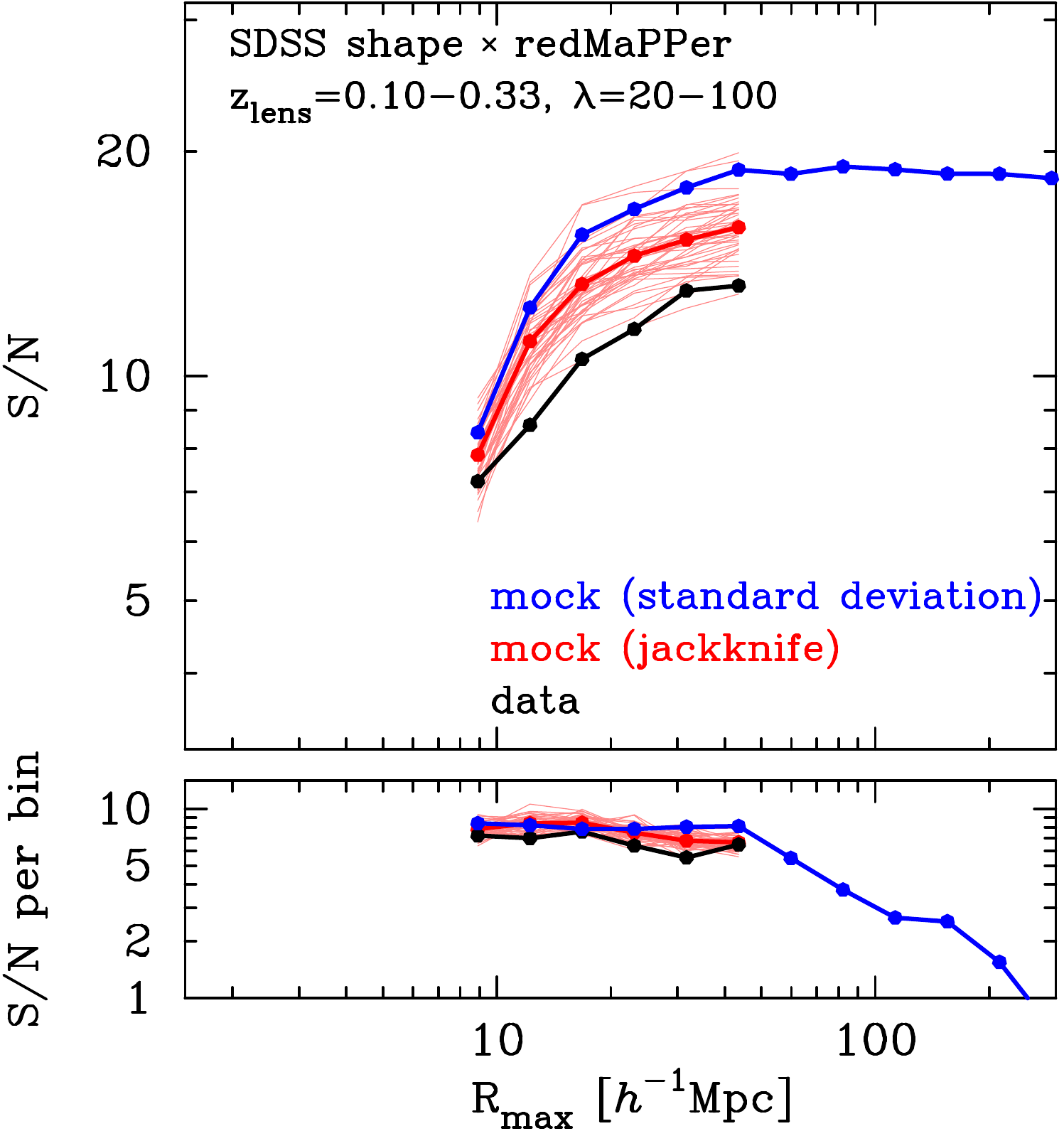}
\includegraphics[width=0.45\columnwidth, bb=0 0 433 458]
{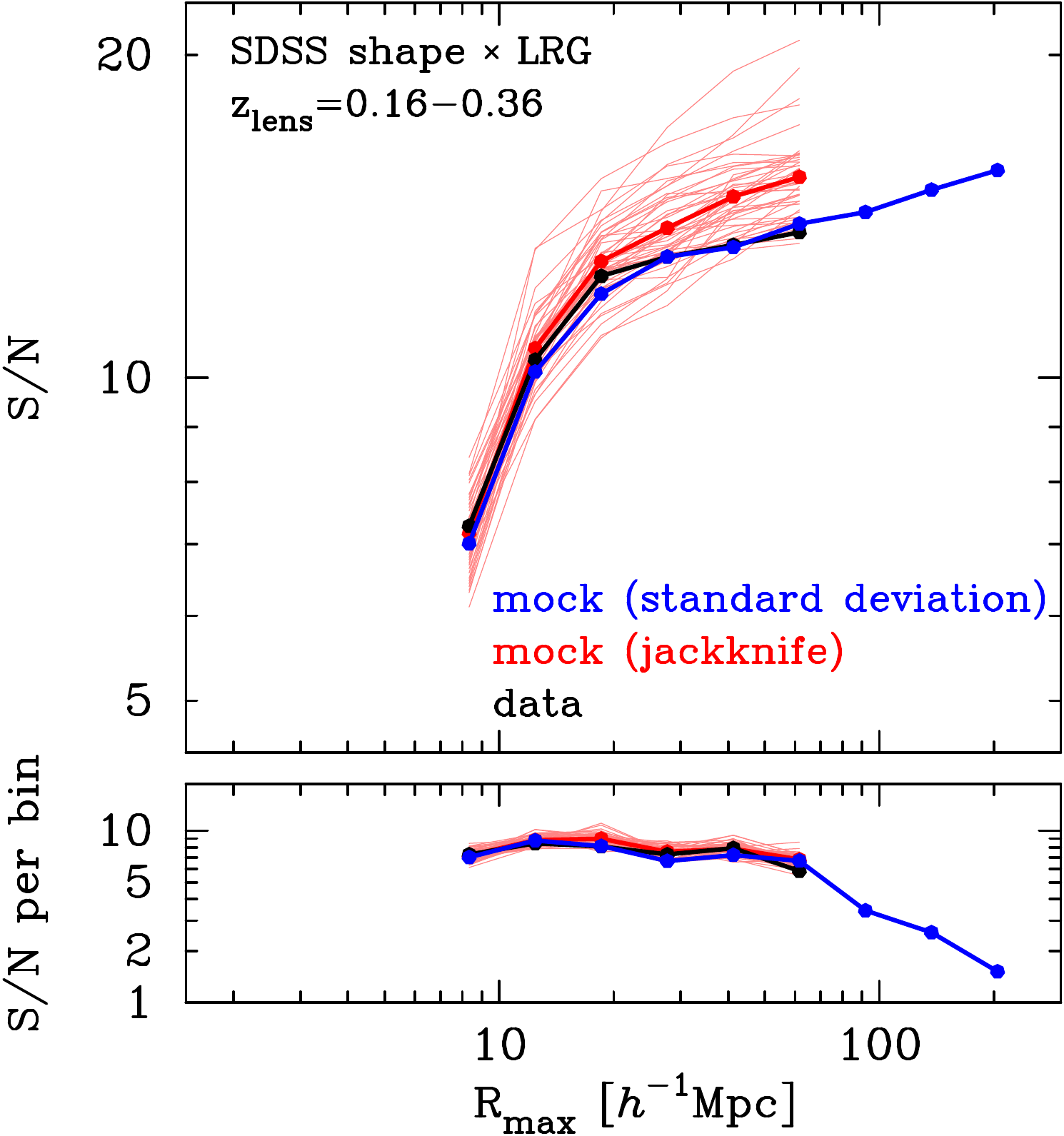}
 \caption{
 {\em Upper panels}: The cumulative signal-to-noise ratio
 ${\rm S/N}$, integrated over $8[h^{-1}\, {\rm Mpc}]\le R\le R_{\rm max}$,
 as a function of $R_{\rm max}$ for the stacked lensing measurements for
 the redMaPPer cluster and LRG samples estimated from the real data and
 the mock catalogs. The stacked lensing signals at $R\simgt 10~
 h^{-1}\, {\rm Mpc}$ are in the two-halo term regime and expected to contain a
 cleaner cosmological information. To compute the curve for the full
 covariance case, we used the actual signals measured from the real data
 up to $R\simeq 50~h^{-1}\, {\rm Mpc}$, and then used the simulated signals at the larger separations.  
 The lower panels show the ${\rm S/N}$ value at each separation bin.
 The use of the full covariance enables us to use larger separation
 beyond the size of JK subregion, where 
 there are signals with ${\rm
 S/N}\simgt 1$ still available in each bin.  } \label{fig:sdss_SN}
\end{figure*} 

In order to quantify the information content obtained from the use of
the accurate covariance, we study the cumulative signal-to-noise ratios
for measuring the stacked lensing in the sample variance limited regime:
\beqa
\left(\frac{{\rm S}}{{\rm N}}\right)^2
=
\sum_{R_{\rm min}<\left\{R_{i},R_j\right\}<R_{\rm max}} 
\widehat{\Delta \Sigma}(R_{i})\, {\bf C}_{ij}^{-1}\, \widehat{\Delta \Sigma}(R_{j}),
\label{eq:def_ourSN}
\eeqa
where $\widehat{\Delta \Sigma}(R_{i})$ denotes the stacked lensing
signal for the $i$-th radius $R_{i}$, ${\bf C}$ is the covariance
matrix, and ${\bf C}^{-1}$ is its inverse. 
When inverting the matrix, 
we use the singular value decomposition for 
the matrix of ${\bf C}$ in the range of $R$ defined by equation~(\ref{eq:def_ourSN}).
Note that, when estimating
the inverse matrix ${\bf C}^{-1}$, we included 
the correction factor of
$N_{\rm r}/[N_{\rm r}-N_{\rm d}-1]$ proposed in \citet{Hartlap2007},
where $N_{\rm r}$ is the number of the realizations and
$N_{\rm d}$ is the dimension of data.  
The summation runs over all the radial bins in the range of $R_{\rm
min}<R<R_{\rm max}$. Here we set $R_{\rm min}\simeq8\, h^{-1} {\rm Mpc}$
in order to focus on the scales in the weakly nonlinear or linear
regime, the so-called two-halo term, where a cleaner cosmological
information can be extracted. 
The cumulative signal-to-noise ratio
includes the contribution of off-diagonal covariance components and is
independent of the radial bin width, while the covariance matrix itself depends
on the width (the Gaussian covariance term depends on the bin width,
since the number of lens-source pairs in a given bin depends on the
width).

Fig.~\ref{fig:sdss_SN} shows the cumulative signal-to-noise ratio,
${\rm S/N}$, for the redMaPPer clusters and LRGs samples as a
function of the maximum separation $R_{\rm max}$.  To compute these, we
use the signals of the real SDSS data up to the JK maximum scale, and
then use the signal measured from the mock catalog at the greater
separations. As for the covariance matrix, we use the different
covariance matrices in Fig.~\ref{fig:sdss_dSigma}, which are estimated
from the real data itself based on the JK method or from the 48
realizations of the mock catalogs based on the full covariance or JK
method.  The thin red curves show the ${\rm S/N}$ values for each of
48 realizations, while the black curve in each panel is the value
obtained from the SDSS data \citep{2016PhRvL.116d1301M}.  Note that the red and black curves are shown up
to $R_{\rm max}\sim50-70\, h^{-1}\, {\rm Mpc}$ corresponding to the size
of JK subregion. The figure shows 
that the ${\rm S/N}$ value for the
real data is on a relatively lower side of the 48 realizations, perhaps
due to the statistical variance.
The bottom
panel of each plot shows the signal-to-noise ratio per each bin,
displaying ${\rm S/N}\ge 1$ for the lensing signals up to the
separation $R\simeq 100~h^{-1}\, {\rm Mpc}$, beyond a size of the JK subregion.  
Such large-scale lensing signals
will be useful to extract useful cosmological information on the
baryonic acoustic oscillations, the parameters of primordial power
spectrum \citep[e.g.,][]{2009PhRvD..80l3527J}, the primordial
non-Gaussianity \citep{2008PhRvD..77l3514D}, and neutrino masses
\citep[e.g.,][]{Takadaetal:06,2009PhRvD..80h3528S,2014PhRvD..90h3530L}.

\section{CONCLUSION and discussion}
\label{sec:conclusions}

In this paper, we have developed a method to create a mock catalog of
the cross-correlation between positions of lensing objects and shapes of
background galaxies -- the stacked lensing of galaxy clusters or
galaxy-galaxy weak lensing. To do this, we fully utilized the full-sky,
light-cone simulations based on a suit of multiple $N$-body simulation
outputs, where the lensing fields of source galaxies are given in
multiple shells in the radial direction out to a maximum source redshift
$z_{\rm s}\simeq 2.4$ as well as the halo catalogs in lens planes are
given in the light cone. Our method enables one to generate a mock catalog
of the stacked lensing following the procedures; (1) define the survey
footprints based on the assigned RA and dec coordinates in the full-sky
simulation, (2) populate the real catalog of source galaxies into the
light-cone simulation realization according to the angular position (RA
and dec) and redshift of each source galaxy, (3) randomly rotate the
ellipticity of each source galaxy to erase the real lensing effect, (4)
simulate the lensing effects on each source galaxy, using the lensing
fields in the light-cone simulation, and (5) identify halos that are
considered to host galaxies or clusters of interest, according to a
prescription to connect the galaxies or clusters to halos (e.g. the
scaling relation between halo mass and cluster richness or the halo
occupation distribution model for galaxies). With this method, we can
use the observed properties of data, the survey footprints and the
positions and characteristics of source galaxies (the intrinsic
ellipticities and the redshift distribution). We applied this method to
the real SDSS catalog of source galaxies as well as the SDSS catalogs of
redMaPPer clusters or luminous red galaxies (LRGs), as shown in
Fig.~\ref{fig:sdss_footprint}. We then showed that our mock catalogs
well reproduce the signals as well as the jackknife (JK) covariance
error bars, estimated from the real data
(Fig.~\ref{fig:sdss_dSigma}). Our method will be powerful to estimate
the error covariance matrix for ongoing and upcoming wide-field weak
lensing surveys.

By having the accurate mock catalogs of cluster/galaxy-shear cross
correlation, we were able to study the nature of the error covariance
matrix.  In particular we focused on addressing validity and limitation
of the JK method, which has been often used in the literature. The JK
method is based on the real data itself, therefore referred to as an
internal covariance estimator, and is known in the fields of statistics
or data science to give an unbiased estimator of the covariance if the
field is Gaussian or Poissonian.  We found that the JK method gives a
reasonably accurate estimation of the true covariance to within 10$\%$
in the amplitude on average, at separation scales smaller than the size
of JK subregion, but it can under-estimate the true error at the larger
separations, especially for a survey with higher number density of
source galaxies as in ongoing and upcoming surveys such as the Subaru
HSC survey.
However we should keep in mind a limitation for the use of the JK
method: the JK covariance can be noisy on each realization basis,
because the JK covariance is estimated from a particular realization
(i.e. data).  The JK covariance matrix becomes noisier or unreliable if
the number of JK subregions/resamples is small or if the dimension of
data vector is comparable with the number of JK subregions.  Thus the
use of accurate covariance matrix for the stacked lensing measurement is
important in the future cosmological analysis.
The full covariance gives us an access to larger separations beyond the
scale of JK subregion, where the JK covariance ceases to be valid
(Fig.~\ref{fig:var_dSigma_Nsub}).
We showed that the full covariance gives 
signals with ${\rm S/N}\ge 1$ out to about $100~h^{-1}\, {\rm Mpc}$ in
the projected separation.  Thus the use of accurate covariance for the
stacked lensing is highly desirable in order to attain the full
potential of ongoing and upcoming surveys. In particular such
large-scale weak lensing signals are expected to contain useful
information on the fundamental physics such as the baryon acoustic
oscillations, the primordial power spectrum, the primordial
non-Gaussianity and the neutrino mass. Exploring an improvement in the
cosmological parameters from the SDSS data by the use of the accurate
covariance is our future work and will be presented elsewhere.

Combining the stacked lensing and auto-correlation
measurements for the same foreground tracers allows one to improve
cosmological constraints by recovering the underlying dark matter
clustering against the bias uncertainty
\citep{Seljaketal:05,2013MNRAS.432.1544M,2015ApJ...806....2M}. This
would also be true even if the foreground objects are affected by the
assembly bias uncertainty if the two measurements are properly combined
\citep{2016PhRvL.116d1301M,McEwenWeinberg:16}. Furthermore it would be
interesting to combine the stacked lensing with the redshift-space
distortion (RSD) measurement for the same foreground galaxies in order
to improve cosmological constraints as well as test gravity theory on
cosmological scales,
by calibrating small-scale systematic effects in the RSD measurement
such as the Finger-of-God effect
\citep{Hikageetal:13,2013MNRAS.435.2345H}. 
A joint experiment of
galaxy weak lensing and CMB weak lensing for the same foreground
clusters/galaxies can be used to directly measure the lensing efficiency
function, $\Sigma_{\rm cr}(z_{\rm l},z_{\rm s})$, without being affected
by nonlinear structure formation including unknown baryonic physics as
well as to calibrate multiplicative systematic biases in the CMB lensing
or galaxy weak lensing that is otherwise difficult to calibrate with
either method alone \citep{DasSpergel:09,Schaanetal:16, 2016arXiv160505337M}. 
Thus once one
starts to combine different clustering measurements, the
dimension of data vector quickly increases and the calibration of auto-
and cross-covariances is more demanding. For this reason, it would be
desirable to develop a hybrid method of combining the mock light-cone catalog of
large-scale structures
containing various fields (weak lensing, halos and velocity
fields) and the analytical method to model the SSC terms in different
observables. This seems feasible, and will be our future work.


\section*{acknowledgments}
We thank Rachel Mandelbaum for making the
SDSS galaxy catalogs and the jackknife subdivisions available to us as
well as for useful discussion.  We thank Nick Battaglia, Bhuvnesh Jain,
Surhud More, Uros Seljak and Matias Zaldarriaga for useful discussion.
M.S. is supported by Research Fellowships of the Japan Society for the
Promotion of Science (JSPS) for Young Scientists.  H.M. is supported by
the Jet Propulsion Laboratory, California Institute of Technology, under
a contract with the National Aeronautics and Space Administration.
Ry.M. is supported by Advanced Leading Graduate Course for Photon
Science.  This work was in part supported by Grant-in-Aid for Scientific
Research from the JSPS Promotion of Science (JP23340061 and JP26610058),
MEXT Grant-in-Aid for Scientific Research on Innovative Areas
(15H05887, JP15H05893, JP15K21733, JP15H05892) and by JSPS Program for Advancing
Strategic International Networks to Accelerate the Circulation of
Talented Researchers.  Numerical computations presented in this paper
were in part carried out on the general-purpose PC farm at Center for
Computational Astrophysics, CfCA, of National Astronomical Observatory
of Japan.

\clearpage

\appendix
 
\section{Dependence of number of subregions on off-diagonal components of the jackknife covariance matrix}
\label{app:off_diag_dSigma_Nsub}

\begin{figure}
\centering
\includegraphics[width=0.52\columnwidth, bb = 0 0 545 504]
{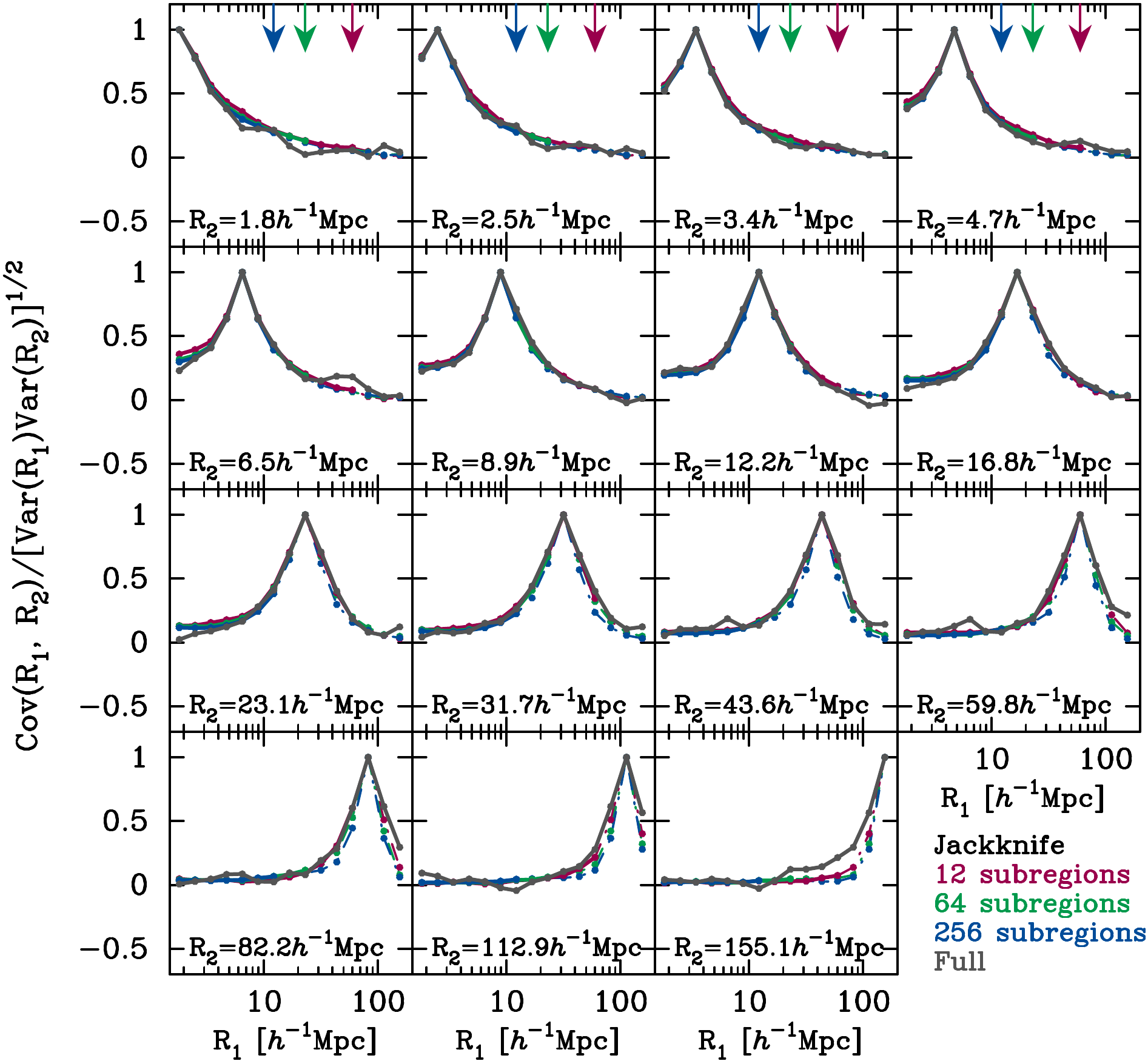}
\caption{
The off-diagonal components of the jackknife covariance
normalized the diagonal components, 
$r_{ij}=C(R_i,R_j)/\sqrt{C(R_i,R_i)C(R_j,R_j)}$, 
estimated from the mock realizations of galaxy-galaxy weak lensing measurements.
We use the same realizations as in Figure~\ref{fig:var_dSigma_Nsub}.
The different panels show the cross-correlation coefficients for different separation $R_i$ as
indicated by the legend.  
In the top panels, the colored arrows show the corresponding 
$R$ to the size of each JK subregion.
}
\label{fig:cov_jk_dSigma_Nsub}
\end{figure}

In this appendix, we show the off-diagonal components of JK covariance
matrices measured from the mock simulations of galaxy-galaxy lensing.
We assume a hypothetical survey with the sky fraction of $f_{\rm sky}=1/48$
and the source redshift of $z_s=0.5$.
As lens objects, we consider a sample of mass-selected halos with
 $M_{\rm 200m}\ge 10^{13.5}\, h^{-1}\, M_\odot$ and 
 in the redshift range $z_{\rm l}=[0.1,0.27]$.
In order to compute the covariance, 
we use the realizations as dubbed 
``S859-full", ``S859-jk1", ``S859-jk2" and ``S859-jk3" in Table~\ref{tab:mass_limited_param}.
Fig.~\ref{fig:cov_jk_dSigma_Nsub} shows the off-diagonal components of JK covariance
when we vary the size of JK subregion.
We found that the impact of the number of JK regions 
on the covariance estimation is more prominent for the diagonal part.
Once the off-diagonal components are normalized by the diagonal components, 
the effect of number of subregions is found to be
$\sim20\%$ at largest.

\section{Area dependence in off-diagonal components of the full covariance matrix}
\label{app:off_diag_dSigma_fsky}

\begin{figure*}
\centering
\includegraphics[width=0.52\columnwidth, bb = 0 0 545 504]
{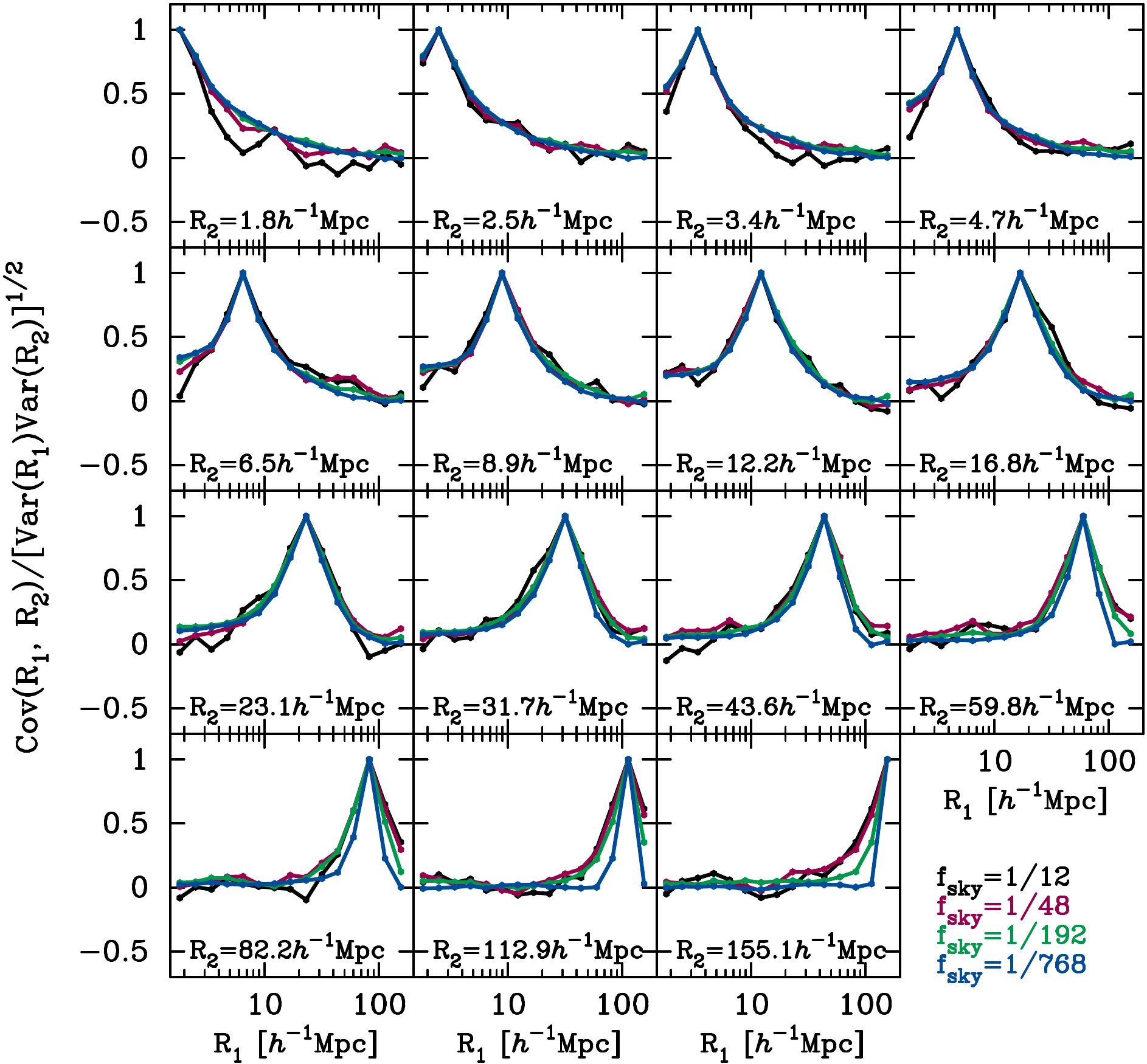}
\caption{
The off-diagonal components of the jackknife covariance
normalized the diagonal components, 
$r_{ij}=C(R_i,R_j)/\sqrt{C(R_i,R_i)C(R_j,R_j)}$, 
estimated from the mock realizations of galaxy-galaxy weak lensing measurements.
We use the same realizations as in Figure~\ref{fig:var_dSigma_fsky}.
}
\label{fig:cov_dSigma_fsky}
\end{figure*}

In this appendix, we examine the area dependence in the off-diagonal components of full covariance
matrices as in Section~\ref{subsec:area_dependence_full_cov}.
Here we use the realizations of ``SXXX-full'' in Table~\ref{tab:mass_limited_param}.
The results are summarized in Fig.~\ref{fig:cov_dSigma_fsky}.
The area dependence in the off-diagonal components is 
found to be relatively small compared to the diagonal one.

\section{Off-diagonal components of the SDSS covariance matrix}
\label{app:off-diagonal}

\begin{figure}
\centering
\includegraphics[width=0.52\columnwidth, bb = 0 0 532 504]
{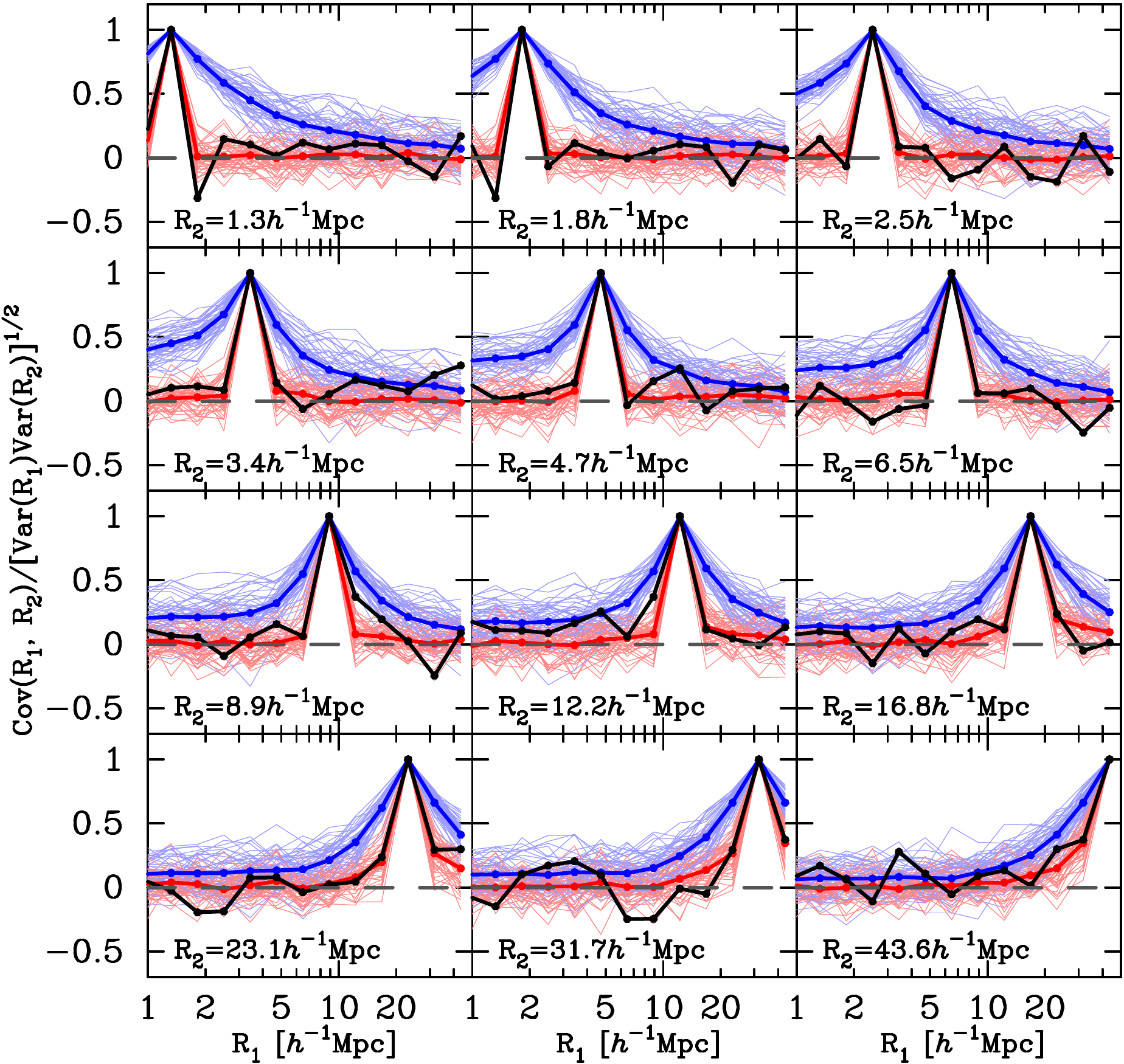}
\caption{The off-diagonal components of the jackknife covariance
 normalized the diagonal components,
 $r_{ij}=C(R_i,R_j)/\sqrt{C(R_i,R_i)C(R_j,R_j)}$, estimated from the real data and
 mock catalogs of the redMaPPer clusters. The different panels show the
 cross-correlation coefficients for different separation $R_i$ as
 indicated by the legend. The black line shows the result from the real
 data, and the thin red line  shows the result obtained from each of the
 48 mock catalogs. The bold red line is the average of the 48
 results. The blue lines are the results for the mock catalogs without
 shape noise. 
	}
\label{fig:jk_mat_redMaPPer}
\end{figure}

\begin{figure}
\centering 
\includegraphics[width=0.52\columnwidth, bb = 0 0 540 508]
{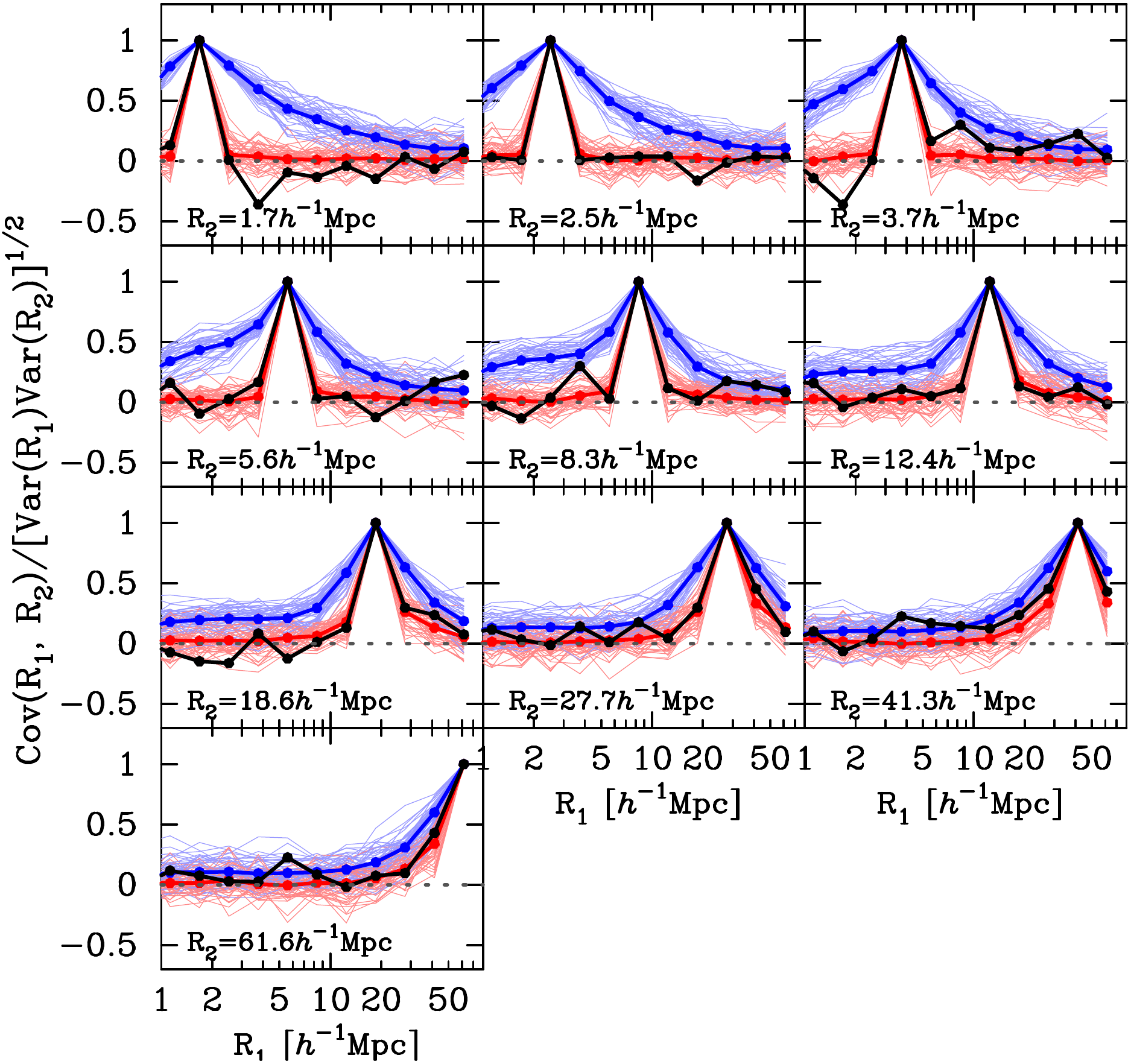}
 \caption{
Similar to
Fig.~\ref{fig:jk_mat_redMaPPer}, but for the luminous red galaxies.  }
\label{fig:jk_mat_LRG}
\end{figure}

In this appendix, we show the off-diagonal components of JK covariance
matrices measured from 48 mock simulations of galaxy-galaxy lensing.
Figs.~\ref{fig:jk_mat_redMaPPer} and \ref{fig:jk_mat_LRG} show the
off-diagonal components for the redMaPPer clusters and LRGs,
respectively, comparing the results from the actual data and the mock
catalogs. The figures show that the JK covariance obtained from one
particular realization can be noisy, and is biased from the underlying
true covariance. Here the average matrix of the 48 JK covariances is
considered to be close to the full covariance for the JK subregion (see
main text for details).
%
%
\clearpage


\bibliographystyle{mnras}
\bibliography{refs_shirasaki}

\label{lastpage}

\end{document}